# Microstructure and Soft Glassy Dynamics of Aqueous Laponite® Dispersion

Khushboo Suman and Yogesh M. Joshi*

*Department of Chemical Engineering, Indian Institute of Technology Kanpur, India*

* Corresponding author, E-mail: joshi@iitk.ac.in

**Abstract**

Synthetic hectorite clay Laponite RD/XLG is composed of disk-shaped nanoparticles that acquire dissimilar charges when suspended in an aqueous media. Owing to their property to spontaneously self-assemble, Laponite is used as a rheology modifier in a variety of commercial water-based products. Particularly, aqueous dispersion of Laponite undergoes liquid - to - solid transition at about 1 volume % concentration. The evolution of the physical properties as dispersion transforms to solid state is reminiscent of physical aging in molecular as well as colloidal glasses. The corresponding soft glassy dynamics of an aqueous Laponite dispersion, including the rheological behavior, has been extensively studied in the literature. In this feature article we take an overview of recent advances in understanding soft glassy dynamics and various efforts taken to understand the peculiar rheological behaviors. Furthermore, the continuously developing microstructure that is responsible for eventual formation of soft solid state that supports its own weight against gravity has also been a topic of intense debate and discussion. Particularly extensive experimental and theoretical studies lead to two types of microstructures for this system: an attractive gel-like or repulsive glass like. We carefully examine and critically analyze the literature and propose a state diagram that suggests aqueous Laponite dispersion to be present in an attractive gel state.

**Introduction**

Colloidal particulate systems in a suspension form are ubiquitous in nature and in industry.[1] Colloidal dispersions have a broad range of applications that depend on the nature of the suspended particles, including their shapes, sizes, charges they possess, and characteristics of the suspending fluid. Particularly the dispersions of particles having dimensions in the nanometer range, and anisotropic shape are of special significance. Owing to surface functionality and extremely high surface area per unit mass, such anisotropic particles spontaneously self-assemble in a liquid media to show a rich array of microstructures.[2] One of the most prominent and widely available anisotropic nanoparticle system is clays that has been studied in the literature over centuries. Clays have plate-like shape with an aspect ratio in a range of 10 to 1000, and possess dissimilar charges in the aqueous media. Among various types of clays, aqueous dispersion of smectite hectorite clay: Laponite RD/XLG® (a registered trademark of BYK Additives) has received significant attention over the past few decades owing to its spectacular physical behavior, which is observed at a relatively low concentration ($<$ 2 volume %).[3-7] On one hand, aqueous dispersion of Laponite is proposed to show a broad spectrum of microstructures, while on the other hand, time and deformation field dependent behavior of the same is considered to be a model to study a variety of industrial pasty systems that show soft glassy dynamics.[8] In this feature article, we discuss various recent developments on understanding phase behavior and soft glassy dynamics of an aqueous dispersion of Laponite RD/XLG.

Clay minerals have a definite crystalline structure which comprises primarily of oxides and hydroxides of different inorganic elements like silicon, aluminium, magnesium, lithium, etc. The clay minerals with sheet like structure belong to the family of phyllosilicates, whose primary building blocks are tetrahedral and octahedral sheets of its constituent elements. The various ways of stacking of these sheets combined with types of elements present and its water absorbing capacity result in a diverse variety of clay minerals.[9] Particularly in 2:1 type of clays two tetrahedral sheets sandwich one octahedral sheet to form a unit cell. Usually, the cations that reside on the outer surface of the tetrahedral sheets dissociate in water rendering it a negative charge.[9-11] The edge charge is due to protonation or deprotonation of the broken crystals depending upon the nature of the aqueous media.[12] It has been proposed that the particles of clay may indulge in a variety of attractive interactions. For a suspension of charged plate-like particles with opposite charges on their faces and edges, the electrostatic attraction may lead to edge-to-face association while attractive van der Waals interaction may drive the face-to-face and edge-to-edge association.[9] The particles, on the other hand, may also experience face to face repulsive interaction.[9] As a result, when dispersed in water, clay particles form mesoscopic structures that have wide applications in different industries. Consequence of very dominant interparticle interactions in clay suspensions was recognized by Bingham[13] over a century ago who studied clay suspension and reported its inability to flow unless a critical finite stress, known as yield stress, is overcome. On the other hand, in a seminal contribution, Langmuir[14] studied the thixotropic behavior of clay suspension and the role of interparticle interactions on the same. Discussion on recent advances in physics, chemistry and applications of clay minerals can be found elsewhere.[11]

Over the past two decades aqueous dispersion of Laponite RD/XLG, a clay mineral belonging to family of phyllosilicates, has generated a lot of interest in the literature.[3, 4, 10, 15, 16] There are many grades of Laponite available in the market, however the grades RD and XLG are chemically identical with latter being purer than the former. In this article, we focus only on these grades. Laponite clay mineral, whose particles are fairly monodispersed, has a diameter around 25-30 nm with thickness around 0.92 nm as shown in figure 1(a).[17] It is typically observed that incorporation of Laponite in water in fairly low concentration causes viscosity and elasticity of dispersion to increase continuously suggesting it to be thermodynamically out of equilibrium.[18] After a certain elapsed time, which strongly depends on the concentration of Laponite, dispersion supports its own weight against gravity. The arrangement of Laponite particles that is responsible for such behavior, particularly above 2 wt. % has been a subject of intense debate in the literature.[3, 4, 6, 7, 10] Regarding the microstructure, there are two schools of thought. One proposes Laponite particles to remain in a self-suspended state stabilized by repulsive interactions among the faces without touching each other, leading to repulsive (Wigner) glass microstructure. The other school proposes Laponite particles to form house of cards structure, wherein edge to face attractive interactions lead to percolated gel like structure. A wide range of characterization techniques has been employed in the literature to investigate the microstructure of Laponite dispersion. Furthermore, the microstructure as well as its evolution as a function of time is very sensitive to temperature and pH of the medium as well as in the presence of external additives such as different kinds of salts and polymers.[19] Interestingly effect of these variables also throws light on the microstructure of neat Laponite dispersion. Moreover, the sample preparation protocol varies amongst different groups, which can noticeably alter the aging behavior of Laponite dispersion hence highlighting the need to set a standard sample preparation protocol.[7] In this article, we critically analyze investigations through different characterization techniques and effect of the above mentioned variables in order to obtain a unified picture of the microstructure.

Moreover, the continuous evolution of viscosity and elasticity of Laponite dispersion indicates progressive slowing down of the microstructural dynamics that is reminiscent of physical aging observed in the structural glasses.[18, 20] The relaxation dynamics of Laponite dispersion that accompanies aging shows a rich array of behaviors. Furthermore, application of deformation field to 'physically aged' dispersion decreases elasticity and viscosity of the same, a behavior that is redolent of rejuvenation that occurs in structural glasses upon increase in temperature.[10, 21] The properties of Laponite dispersion strongly depend on competition between the nature of structural evolution that occurs during aging and the breakdown that occurs during rejuvenation leading to a spectrum of complex rheological behaviors. In this article, we discuss the physical origins of such behaviors and the corresponding modelling efforts. Furthermore, by virtue of inherent time dependency, the aqueous dispersion of Laponite does not obey the fundamental principles of viscoelasticity. The experimental and modelling efforts that render the laws of viscoelasticity applicable to the Laponite dispersion are also reviewed in this article.

When dispersed in liquid the faces of Laponite disks acquire a negative charge[9] and the dissociated $Na^+$ counter-ions form an electric double layer on the same. Consequently, since the interparticle interactions among the Laponite particles in aqueous media are governed by electrostatic and van der Waals potentials, Laponite dispersion becomes an ideal system to apply DLVO theory and test its predictions by analyzing the microstructural evolution.[10, 22] Owing to its remarkable rheological behavior, Laponite has been used in numerous products as a rheology modifier in aqueous systems. Its applications span a very diverse set of industries from pharmaceutical, home and personal care, agrochemical, petrochemical, petroleum to construction. Lately, Laponite dispersed in aqueous media has also been used for various novel applications such as carrier for drug delivery[23], glucose biosensors[24, 25], orthopaedic tissue repair[23], bioactive materials,[23, 26] electrodes,[25, 27] etc.

An outline of this article is as follows. We first begin with the introduction of Laponite particle and then in the following section, we discuss the characteristics of an aqueous dispersion of Laponite. Subsequently, we focus on the chemical stability of Laponite dispersion and then present soft glassy dynamic characteristics of Laponite dispersion. We then review the investigation of microstructural evolution in Laponite dispersion via different characterization techniques. In the next section, we propose a mechanism for microstructural evolution and a state diagram. Finally, we present the open problems and future directions followed by the concluding remarks.

**Laponite particle:**

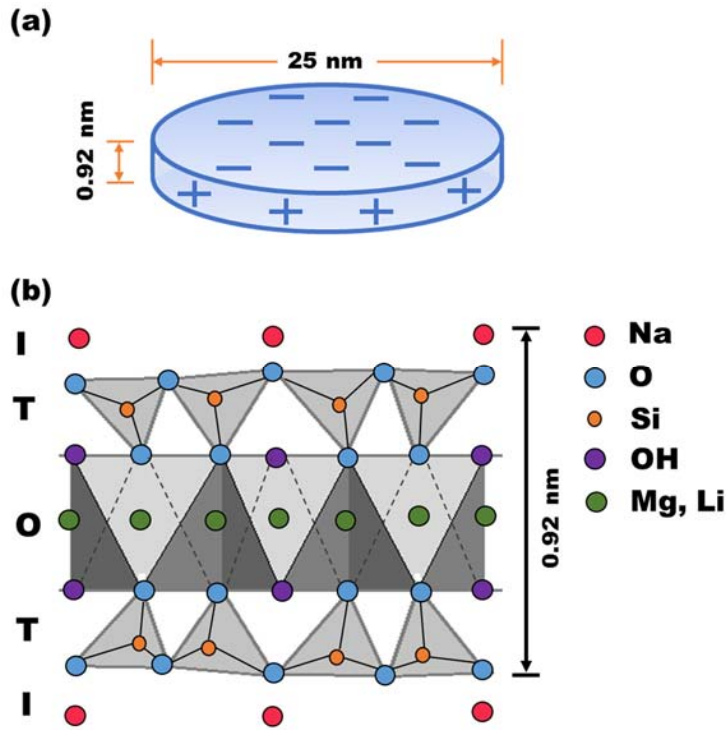

**Figure 1.** (a) Schematic of a single particle of Laponite in an aqueous dispersion at pH=10 (b) Idealized atomic arrangement in the unit cell of Laponite clay comprising of the octahedral sheet (O) sandwiched between two tetrahedral sheets (T). The 2:1 layer coordinates with interlayer (I) gallery of sodium atoms.

Hydrous sodium lithium magnesium silicate, commonly known as Laponite, is a 2:1 synthetic hectorite clay mineral. The building blocks of a Laponite particle are two-dimensional array of silicon-oxygen tetrahedra that is sandwiched by arrays of oxygen-magnesium-hydroxyl octahedra on either side. The side view of an atomic arrangement in a Laponite particle is shown in Fig 1(b). Within each layer, a unit cell with chemical formula: $Na_{0.7}[(Si_8Mg_{5.5}Li_{0.3})O_{20}(OH)_4]$, which is depicted in figure 1(b), repeats itself in the lateral direction.[9] The dimensions of the unit cell, 5.15Å ×8.9 Å ×9.2 Å,[9, 28, 29] leads to an approximate face area of a unit cell to be 0.458 nm$^2$. In the literature, the diameter of Laponite particle has been reported to be 25±0.25 nm[17] and the molecular weight to be $7.1\times10^5 \sim 9.3\times10^5$ g/mol,[30, 31] which suggest that on an average 1100±100 unit cells constitute a single particle of Laponite. Considering industrial production of Laponite a large size distribution is expected, and some reports do claim the diameter to be in a range 20 to 30 nm.[30, 32] The isomorphous substitution of divalent magnesium by monovalent lithium (occurring in octahedral sheet) and presence of vacant sites results in deficit of positive charge within the cell, which renders a net negative charge to the faces of a disk.[9, 11] This negative charge is compensated by positively charged sodium ions that reside in the interlayer gallery. Considering 1000 unit cells, around 700 sodium atoms are associated with a particle,[7] that are located on both the sides of a disk. In dry form, Laponite particles are stacked parallel to each other and are termed as tactoids. The aggregates of such particle tactoids have a typical macroscopic dimension in a range of 1 to 50 μm. Dry powder of Laponite is highly hygroscopic and can pick up to 20 mass % of moisture when exposed to humid conditions. It is, therefore, necessary to properly dry the powder before using it for preparation of aqueous dispersion. In a dry state, the bulk density of Laponite powder is 1000 kg/m$^3$ while the particle density is 2530 kg/m$^3$.

**Characteristics of aqueous dispersion of Laponite**

Laponite in its raw state is primarily used in the form of aqueous dispersion. In the presence of water, the sodium cations on the outer faces of a particle can be exchanged with other cations. The cation exchange capacity (CEC) of Laponite particle has been reported in the range 0.55-0.8 meq/gm of dry Laponite,[32, 33] and is known to decrease with increase in pH.[32] Such value of CEC corresponds to around 450 exchangeable $Na^+$ ions per Laponite particle. On dispersing Laponite powder in water, the tactoids are proposed to undergo interlayer swelling in two steps. In the first step, water molecules penetrate the interlayer gallery that increases the basal spacing and push the disks apart allowing more water to enter. Owing to the surrounding aqueous medium, sodium ions dissociate from the Laponite surface and diffuse into the bulk due to osmotic gradient rendering a permanent negative charge to the faces. This negative charge leads to repulsion between the faces of Laponite particles.[34] In the second step, particles move further apart due to the double layer repulsion.[34] The Zeta potential of Laponite dispersion is reported to be around -40 mV.[35] While charge on the edge of a particle depends on pH of the aqueous media, the faces of a Laponite particle acquire a permanent negative charge. In aqueous media, the edge of Laponite is composed of – Mg–OH groups in the middle and silica groups on the sides. The isoelectric point of –Mg-OH is around 12 while that of silica is around 2.[36] Interestingly, the isoelectric point of the edge of Laponite particle has been observed to be significantly skewed towards that of –Mg-OH and has been reported to be between 10 and 11.[37]

When Laponite powder is dispersed in water it forms a hazy dispersion. The continuous stirring for around 30 min, however, completely clears the same leading to transparent dispersion. It has been observed that incorporation of 2.8 wt. % Laponite in water having initial pH in a range between 3 to 10 changes pH of water to around 10.4.[38] Such increase in pH occurs due to protonation (dissociation of $OH^-$ ions) of the edges of the Laponite particles rendering the edges a positive charge. If the initial pH is above 11, dissociation of $H^+$ ions from the edge leads to a negative charge on the same. If the initial pH is between 10 and 11, the nature of edge charge depends on the concentration of Laponite $\left(C_L\right)$. By knowing the change in pH and number density of Laponite particles $\left(n_p\right)$ in the aqueous media, the average charge on the edge of a single

Laponite particle can be easily estimated by using $E_C = eN_A \left(10^{\text{pH}_f} - 10^{\text{pH}_i}\right)10^{-14}/n_p$, where $e$ is electron charge, $N_A$ is the Avogadro number while $\text{pH}_i$ and $\text{pH}_f$ are respectively the initial and the final of pH of a dispersion. Tawari et al.[12] reported edge charge to be $+50e$ for the dispersion with $\text{pH}_f$ of 9.97 at 25°C. Since the $\text{pH}_f$ for a given concentration has been observed to be almost independent of $\text{pH}_i$, lower value of $\text{pH}_i$ leads to greater intensity of positive charge on the edge.

**Stability of Laponite:**

It has long been debated in the literature that acidic environment leads to the disintegration of hectorite clays in general and Laponite clay in particular.[39] The very first investigation on the stability of Laponite dispersion has been carried out by Thompson and Butterworth[32] in low concentration regime ($C_L < 2$ wt. %). For pH less than 9, they observed degradation of Laponite particles leading to leaching of magnesium ions $\left(\text{Mg}^{2+}\right)$ according to the reaction:[32]

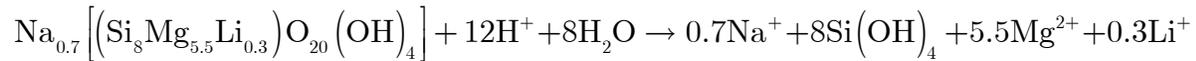

$$\text{Na}_{0.7}\left[\left(\text{Si}_8\text{Mg}_{5.5}\text{Li}_{0.3}\right)\text{O}_{20}\left(\text{OH}\right)_4\right] + 12\text{H}^+ + 8\text{H}_2\text{O} \rightarrow 0.7\text{Na}^+ + 8\text{Si}\left(\text{OH}\right)_4 + 5.5\text{Mg}^{2+} + 0.3\text{Li}^+$$

Further investigation on the chemical stability of Laponite dispersion has been performed by Mourchid and Levitz.[40] They study dynamics of gelation at low concentration ($C_L = 1$ and 1.5 wt. %), wherein they observe that samples stored in air showed traces of $\text{Mg}^{2+}$ whereas samples preserved in nitrogen atmosphere did not. This difference has been attributed to dissolution of $\text{CO}_2$ from air which creates an acidic environment required for $\text{Mg}^{2+}$ leaching. They claim that when pH of dispersions decreases below 9, Laponite particles are prone to chemical instability due to $\text{Mg}^{2+}$ leaching. Both these studies lead to a protocol of preparing Laponite dispersion in water maintained at a higher pH (around 10). Recently, a systematic study on the chemical stability of Laponite dispersion (1, 1.7 and 2.8 wt%) with and without salt, prepared with different initial pH of water (from pH 3 to 10) has been carried out by Jatav and Joshi.[38] They observe leaching of $\text{Mg}^{2+}$ in low concentration dispersion but not in high concentration system, thereby contradicting the belief of $\text{Mg}^{2+}$ leaching occurring only below final pH of 9. The stability in high concentration sample against the attack of $\text{H}^+$ ions is attributed to the greater concentration of $\text{Na}^+$ ions, which could be because of dissociated Na$^+$ counter-ions or from externally added salt. A detailed study of the chemical stability of Laponite dispersion has been reported by Mohanty and Joshi.[41] They examine the stability over a wide range of Laponite ($1 < C_L < 4$ wt. %) and NaCl concentrations ($0 < C_S < 7$ mM) against chemical degradation and propose a chemical stability phase diagram shown in figure 2 indicating the stable and unstable region for different values of $C_L$ and $C_S$ as a function of time. It can be seen that higher is the value of $C_L$ and $C_S$, greater is the chemical stability of dispersion. It has been observed that the leached Mg$^{2+}$ ions enhance the ionic strength, which leads to accelerated aging of Laponite dispersion.[66] Mohanty et al.[42] examine the effect of leaching of Mg$^{2+}$ ions on macroscopic properties using rheology and DLVO theory.

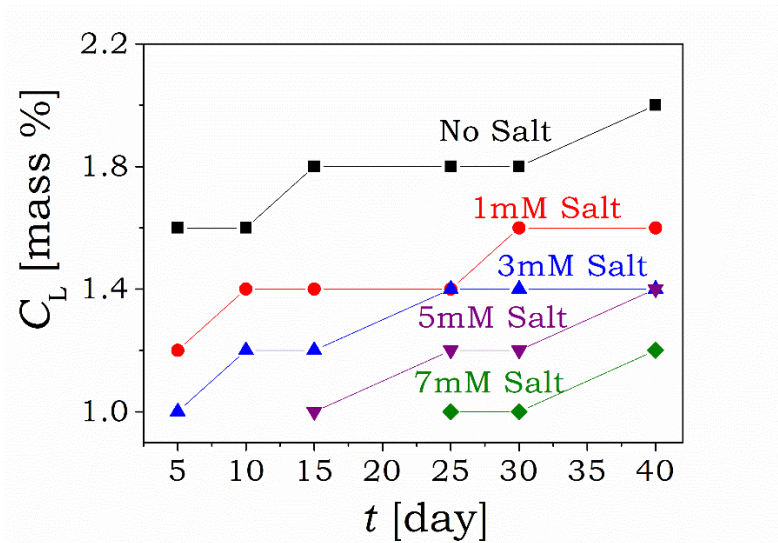

**Figure 2.** Proposed chemical stability phase diagram by Mohanty and Joshi[41] for different values of $C_S$. The region below a line denotes the unstable regime where Mg$^{2+}$ leaching is observed. The region above a line indicates a chemically stable region. Reproduced with permission from reference 41. Copyright 2016 Elsevier.

**Soft Glassy Dynamics**

Aqueous dispersion of Laponite undergoes liquid-solid transition soon after its preparation, wherein its viscosity, modulus and therefore relaxation time continuously increases with time.[19, 43] On the other hand application of stress causes these properties to decrease with time.[43] Both these behaviors are reminiscent of respectively the physical aging and the rejuvenation observed in the glassy systems (including molecular and spin glasses) in general and colloidal glasses in particular.[43-45] Interestingly many soft materials of common interest such as concentrated suspensions and emulsions, colloidal gels, foams, etc., show aging and rejuvenation behavior very similar to that observed for molecular as well as colloidal glasses. As a result, these materials are called as soft glassy materials in the literature.[46] Many materials of commercial importance that have paste-like consistency show soft glassy dynamics. Furthermore, Laponite is added in many water-based industrial systems to render it pasty consistency, which leads them to exhibit soft glassy characteristics. As a result, soft glassy dynamics of aqueous dispersion of Laponite has been investigated as a model system to understand wide class of industrial pasty systems as well as other model soft glassy materials.[10, 43, 47] Furthermore, owing to accessible timescales and length-scales in Laponite dispersion compared to that of in the molecular glassy systems, the former is analyzed to obtain insight into comparatively inaccessible dynamics of the latter.

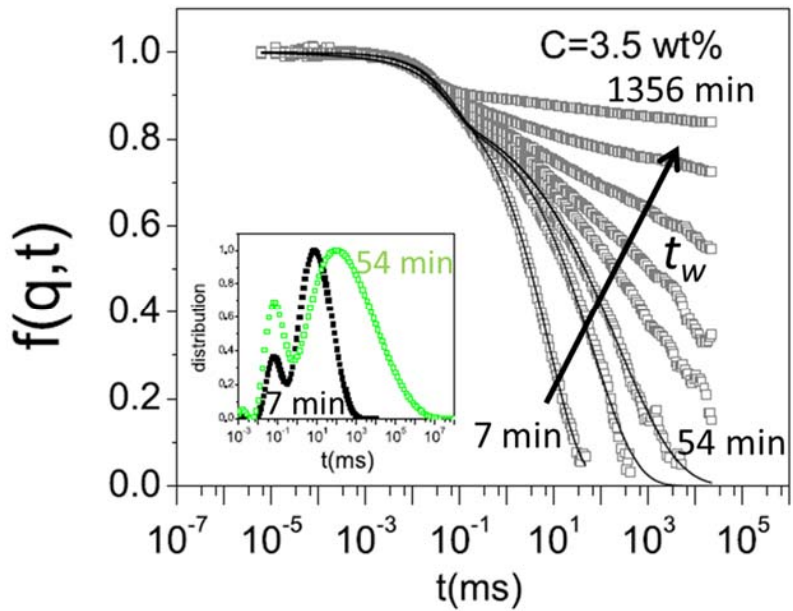

**Figure 3.** Evolution of intermediate scattering function $f(q,t)$ as a function of delay time and corresponding fits to eq (1) for $C_L = 3.5$ wt. %. The waiting time $t_w$ increases from left to right as indicated by arrow. In the inset relaxation time distribution associated with the fast and slow mode is also plotted. The figure shows that while distribution associated with fast mode does not age, the distribution associated slow

mode undergoes aging. Reproduced with permission from reference 48. Copyright 2007 American Physical Society.

Dynamic Light Scattering (DLS) has been used very extensively to study soft glassy dynamics of an aqueous dispersion of Laponite. In DLS study of $C_L > 2$ wt. %, the autocorrelation function $(g_2 - 1)$ has been observed to show a two-step decay as shown in figure 3 from which a fast $(\tau_1)$ and slow $(\tau_2)$ mode is decoupled by fitting:

$$\sqrt{g_2 - 1} = \phi \exp(-t/\tau_1) + (1-\phi)\exp(-(t/\tau_2)^\beta), \tag{1}$$

where $\phi$ and $\beta$ are the fitting parameters.[43, 49-53] In figure 3 we show the decay of $\sqrt{g_2 - 1}$ for $C_L = 3.5$ wt. %.[48] It can be seen that eq (1) fits the data well leading to the estimation of fast and slow timescales. Many glass as well as gel forming systems are known to show two-step decay. For glass forming systems the fast time scale represents rattling motion within a cage while slow timescale represents cage diffusion process as particle hops between the cages.[54] In gels, if $\varsigma$ is the correlation length (of the order of radius of gyration of an aggregate) and $q$ is the wave vector or the inverse of the length-scale probed the fast mode has been attributed to the cooperative diffusion for $q\varsigma < 1$. For $q\varsigma \gg 1$ the fast mode has been assigned to internal modes of individual fractal aggregates [55]. The slow mode indicates diffusion of aggregates as well as the restructuring of a gel.[56] The fast relaxation mode $(\tau_1)$ for Laponite dispersion is observed to be independent of aging time $t_w$ is some studies[47, 57] while it is observed to increase with $t_w$ in other studies.[54, 58-60] Furthermore, $\tau_1$ is observed to show diffusive behavior given by: $\tau_1 \sim q^{-2}$ dependence.[47, 51, 57, 61, 62] Interestingly as shown in figure 4, the slow relaxation mode $(\tau_2)$ has been observed to show an exponential dependence on $t_w$: $\ln \tau_2 \sim t_w$ for a certain duration after preparation of sample, which eventually changes to power law dependence: $\tau_2 \sim t_w^\mu$ with $\mu = 1$ to 2.[49, 50, 57, 61, 63] The regime where exponential dependence emerges is ergodic in nature with $\tau_2 \sim q^{-2}$ and has been termed as cage forming regime.[34, 63] During the exponential regime $g_2 - 1$ has been observed to show an exponential decay ($\beta = 1$) for the slow mode.[57] The subsequent power law regime has been observed to be non-ergodic and has been termed as full aging regime.[50, 57, 63] In this regime some groups observe $g_2-1$ to show compressed exponential decay ($\beta=1.3$ to 1.45) with sub-diffusive dynamics $\tau_2 \sim q^{-x}$ with $x=1$ to 1.3.[50, 57, 64] Conversely for the same regime some groups report stretched exponential dependence ($\beta<1$) with $\tau_1 \sim q^{-2}$ as well as $\tau_2 \sim q^{-2}$ scaling.[47] Interestingly Ianni et al.[58] observe that shear melting the sample before gelation does not affect the relaxation behavior of the same. However, shear melting the sample after gelation causes autocorrelation function to show compressed exponential behavior.

One of the important features of the systems showing glassy dynamics is dynamical heterogeneity.[65-67] Particularly as a system approaches glass transition or structural arrest the dynamics of the constituents become more and more correlated leading to the growth of dynamic correlation length-scale.[68] It causes broadening of the relaxation time distribution, and as a result, the relaxation dynamics becomes strongly non-exponential.[69] The dynamic heterogeneity has also been observed in colloidal glassy systems[65, 67-69] in general and aqueous Laponite dispersion[70-75] in particular. One of the strong indications of dynamical heterogeneity is the non-exponential behavior of a relaxation process or the intermediate scattering function.[66, 76, 77] As discussed above the stretched exponential relaxation dynamics with decrease in parameter $\beta$ with aging time suggests broadening of the relaxation time distribution, which can be attributed to the dynamical heterogeneities and their growth.[52, 78, 79] The size or length-scales of the dynamical heterogeneities are extracted using higher order correlation functions which are the fluctuations of two-point correlation function in space and time or with respect to the change in control parameter. The rigorous mathematical treatment of high order correlation functions that relate the size of dynamical heterogeneities can be found elsewhere.[80, 81] Recently Jabbari et al.[70] investigate the correlation function of aqueous Laponite dispersion and observed the rotational diffusion of particles to decline rapidly than their translational motion in 3 wt. % Laponite dispersion. They attribute this decoupling of rotational and translational relaxation times to development of heterogeneous dynamics. Maggi et al.[71] develop a novel experimental setup, which combined the homodyne and heterodyne DLS technique to study the spatial fluctuations in 1.1 wt. % Laponite dispersion. The authors observe a deviation from Gaussian approximation in the four-point dynamic susceptibility with increasing aging times. The peak in four-point susceptibility has been associated with the amount of dynamically correlated colloidal particles which grows with aging time. Gadige et al.[72] introduce a three-point dynamic susceptibility to study dynamical heterogeneity in Laponite dispersions $\left(2 \leq C_L \leq 3.5\right)$. This three-point dynamic susceptibility has been computed by differentiating the intermediate scattering function with respect to aging time and concentration. Similar to results observed by Maggi et al.,[71] three-point dynamic susceptibility exhibits a peak which increased with aging time and $C_L$. The number of dynamically correlated particles displays a power law dependence on the aging time of Laponite dispersion. Oppong et al.[73] study the distribution of particles in 1 wt. % Laponite dispersion with 1mM salt using microrheological tools. They observe the distribution to deviate from Gaussian with an increase in aging time showing that, on the length scale probed by tracer particles, the dispersion is spatially heterogeneous. This deviation is quantified in terms of a non-Gaussian parameter that illustrates that degree of heterogeneity in the sample increases with aging time. Similar microrheological measurements have also displayed the size-dependent spatial heterogeneities.[74, 75]

Laponite dispersion has been widely studied in the literature to examine the validity of the Fluctuation-Dissipation theorem (FDT). FDT can be corroborated using microrheology by measuring position fluctuations and response function simultaneously, which leads to computation of effective temperature $T_{\text{eff}}$ of a system. In case of $T_{\text{eff}} = T_{\text{bath}}$ (bath temperature), FDT gets validated. Abou and Gallet[82] observe a non-monotonic dependence of $T_{\text{eff}}$ on waiting time while Greinert et al.[83] observe $T_{\text{eff}}$ to increase with time leading to invalidation of FDT. Conversely, Jabbari-Farouji et al.[84, 85] and Ciliberto and coworkers[86] observe a close match between the active (response function) and passive (thermal fluctuations) microrheology results leading to $T_{\text{eff}} \approx T_{\text{bath}}$. Such conflicting results can be attributed to the different aging times and frequencies employed by different groups.

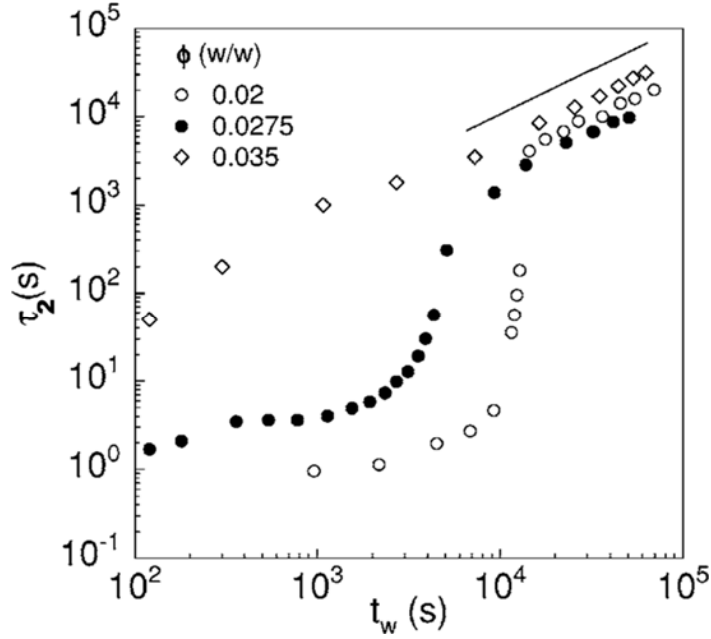

**Figure 4.** Evolution of $\tau_2$ as a function of aging time $(t_w)$. With increase in $t_w$, the dependence of $\tau_2$ changes from exponential (cage forming) to power law (full aging) regime. Reproduced with permission from reference 49. Copyright 2006 American Physical Society.

Soft glassy dynamics of Laponite dispersion has been extensively studied using rheology. Usually, a Laponite dispersion is shear melted before carrying out any rheological experiments. This, in principle, ensures that a sample does not have any prior shear history. In order to explore physical aging, subsequent to shear melting, a material is subjected to sinusoidal strain $\gamma = \gamma_0 \sin \omega t$, where $\gamma_0$ is the strain amplitude in a linear viscoelastic limit, $\omega$ is the frequency and $t$ is the time. A rheometer measures output in terms of stress: $\sigma = \sigma_0 \sin(\omega t + \delta)$, that lags the strain input by angle $\delta$. The knowledge of $\sigma_0$, $\gamma_0$ and $\delta$ for an applied $\omega$ facilitates estimation of elastic $\left(G'(\omega,t) = (\sigma_0/\gamma_0)\cos\delta\right)$ and viscous $\left(G''(\omega,t) = (\sigma_0/\gamma_0)\sin\delta\right)$ moduli. For Laponite dispersion $G'$ and $G''$ usually evolves in two steps as shown in figure 5.[87] In the first step both the moduli increase with time wherein $G'$ increases more rapidly than $G''$ and crosses the same. In the second step increase in $G'$ becomes weak while the $G''$ starts decreasing after showing a maximum. The increase in $G'$ during aging is usually observed in the systems with strong enthalpic contributions.[88, 89] Figure 5 also shows the effect of temperature $(T)$ on the evolution of $G'$ and $G''$. It can be seen that evolution gets accelerated at higher $T$ and shifts to lower times without changing its shape. Interestingly in molecular glasses, physical aging has also been observed to get accelerated with an increase in $T$. Moreover, it has been observed that molecular glasses follow an asymmetric path to equilibration upon temperature step-up and step-down jump.[90] Likewise, Laponite dispersion has also been observed to follow different routes as it evolves subsequent to step-up and step-down temperature jumps.[18]

In a typical experimental protocol, shear melted Laponite dispersion is aged for a certain duration (also known as waiting time or $t_w$) before applying any deformation field.[20, 21, 91] In a typical stress relaxation experiment, similar to DLS, stress is also observed to relax in two steps.[54] The corresponding fast relaxation mode has been observed to show a weak dependence on $t_w$. The slow relaxation mode $(\tau_2)$, obtained by fitting the stress relaxation modulus to $G(t) = \phi \exp\left(-(t/\tau_2)^\beta\right)$, shows a strong dependence on $t_w$ with $\beta \leq 1$.[54] It has usually been observed that stress relaxation in Laponite dispersion becomes exceedingly sluggish as time passes and eventually reaches a plateau associated with the residual stress.[88] Negi and Osuji[92] observed that greater the strength of the flow field (stress or strain rate) during rejuvenation, more is the relaxation in a material and therefore lower is the magnitude of residual stresses.

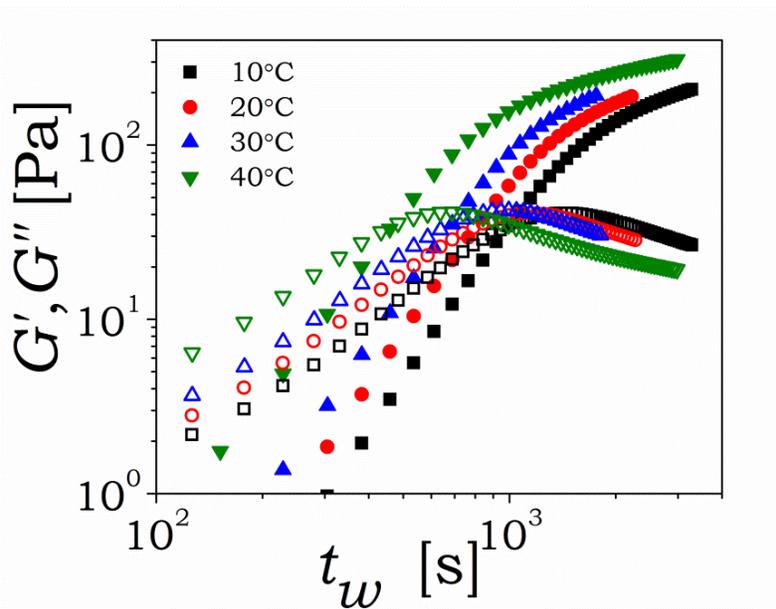

**Figure 5.** Temperature dependent evolution of $G'$ (solid symbols) and $G''$ (open symbols) as a function of time in 2.8 wt. % Laponite dispersion for rest time of 18 days. Reproduced with permission from reference 10. Copyright 2012 American Chemical Society.

Many groups have also performed creep experiments on Laponite dispersion.[88, 91, 93] The viscoelastic nature of Laponite dispersion coupled with the instrument inertia has been observed to give rise to oscillations in a creep experiment, which decay over a period of 1s,[94] lead to an estimation of high frequency elastic modulus.[95] As mentioned before, under quiescent conditions, the viscosity of Laponite dispersion undergoes continuous increase as a function of time. Under application of constant creep stress below a threshold value, the rate of increase in viscosity weakens with an increase in stress but does not stop. Under such conditions, viscous strain indeed gets induced in the material. However, as viscosity diverges strain reaches a plateau.[43] On the other hand, if applied creep stress is above the threshold, viscosity decreases with time and eventually reaches a constant value. The strain in such case continues to increase, and in a limit of large times, shear rate reaches a steady value. In the literature, this behavior has been termed as viscosity bifurcation and the threshold stress at the point of bifurcation has been represented as yield stress.[43] Although this yield stress does not follow the conventional definition of yield stress that forbids viscous flow below it.[89] Very interestingly, the point at which Laponite dispersion yields has also been observed to be dependent not just on applied stress but also on the time of application of stress. It has been reported that application of stress leads to an almost constant value of strain over a long time before it undergoes a sudden yielding.[96] Such delayed yielding has been attributed to gradual alteration of relaxation time distribution of Laponite dispersion under application of stress. In an apparently opposite observation, Laponite dispersion has been observed to undergo delayed solidification, wherein dispersion in a liquid state under application of constant stress or oscillatory stress undergoes sudden solidification.[92, 97, 98] This behavior has been explained by the slow increase in the population of higher relaxation modes leading to the observed behavior.

Many groups obtained the yield stress of Laponite dispersion as a function of $t_w$ by using conventional procedures such as subjecting the material to increasing stress or strain ramp in linear or oscillatory mode. In figure 6 we plot yield stress as well as yield strain data from the three groups as a function of $t_w$.[73, 99, 100] Interestingly it can be seen that yield stress of Laponite dispersion first shows a constant value up to a certain $t_w$ followed by an increase. The corresponding yield strain, on the other hand, initially decreases with time over the duration for which yield stress is constant, and then remains constant.

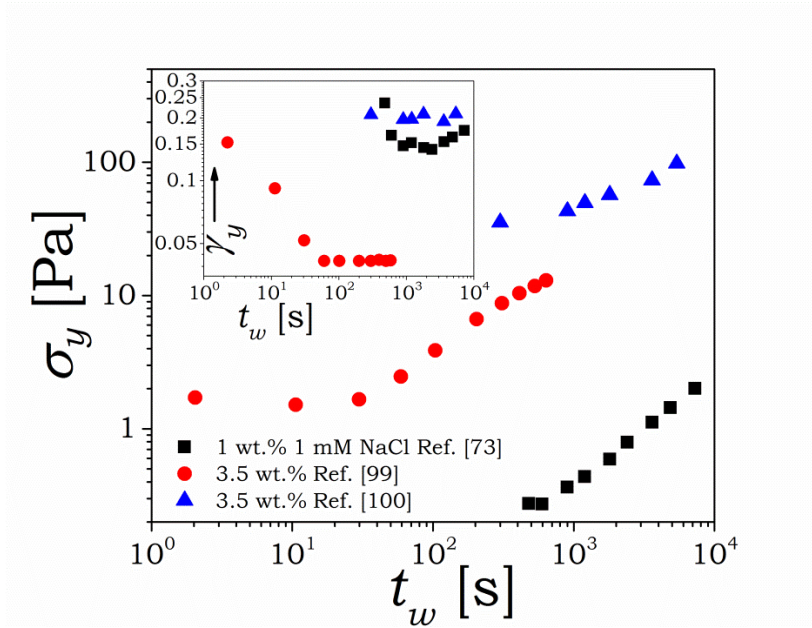

**Figure 6.** Temporal evolution of yield stress $(\sigma_y)$. The inset shows the time evolution of yield strain $(\gamma_y)$. Data from references 73, 99 and 100.

Interestingly, aqueous dispersion of Laponite has been observed to demonstrate transient as well as steady state shear banding in a simple shear flow, where the gradient in stress in a shear cell is very weak. Martin and Hu[101] observe that when the applied constant shear rate is small, Laponite dispersion shows transient shear banding wherein the no-flow band coexists with the flowing band. As time passes the flowing band shrinks and eventually shows steady state shear banding. At moderately high shear rates, interestingly the no-flow band shrinks ultimately leading to steady state shear band. However, for high enough shear rates, the no-flow band shrinks completely to show the homogeneous flow.[101] Recently Jain and coworkers[102] have solved a thixotropic model that shows non-monotonic flow curve in a rectilinear couette geometry by accounting for inertia. Interestingly the model captured all the above-mentioned features of shear banding reported by Martin and Hu[101] very well. Gibaud et al.[103] study shear banding behavior of Laponite dispersion in a shear cell with smooth as well as rough walls. They observe that while smooth wall that promoted wall slip and induced fragmentation of the sample, eventually produced homogeneous flow. The rough walls, on the other hand, prevented wall slip but led to steady state shear banding.

Rheological behavior of any soft material is studied using principles of linear viscoelasticity. The most fundamental principle of linear viscoelasticity is the Boltzmann superposition principle (BSP) that suggests the total response (stress $(\sigma)$ or strain $(\gamma)$) to be a sum of individual responses associated with an independent application of impetus (strain or stress). Generalization of this principle in an integral form is given by:[88]

$$\gamma(t) = \int_{-\infty}^{t} J\left(t - t_w\right) \dot{\sigma} dt_w \text{ and } \sigma(t) = \int_{-\infty}^{t} G\left(t - t_w\right) \dot{\gamma} dt_w, \qquad (2)$$

where $t$ is the present time and $G$ and $J$ are the stress relaxation modulus and creep compliance respectively. Combination of both the expressions lead to convolution relation:[88] $t = \int_{0}^{t} G(s) J(t-s) ds$. However, validation of this principle needs the response functions: $G$ and $J$ to depend only on time elapsed since application of deformation field $(t - t_w)$. For materials that show time dependent change in properties such as Laponite dispersion, $G$ and $J$ show independent dependence on time at which the deformation field is applied $(t_w)$ leading to: $J = J(t - t_w, t_w)$ and $G = G(t - t_w, t_w)$ .[104, 105] As a result, BSP as well as the convolution relation cannot be applied in its conventional form. Furthermore, the invalidity of BSP also does not validate the time-temperature superposition in the real time domain. This poses a significant challenge in applying principles of linear viscoelasticity to the time dependent soft materials in general and Laponite dispersion in particular. To address this issue, Joshi and coworkers[20, 96, 105, 106] through series of papers employ an effective time domain approach for time dependent soft materials. An effective time scale $\xi(t)$ given by:

$$\xi(t) = \tau_0 \int_0^t dt'/\tau(t'),  \qquad (3)$$

is proposed, where $\tau = \tau(t)$ represents the dependence of relaxation time on the real time. The normalization of real time by the time dependent relaxation time as expressed in eq (3) fixes the relaxation time in the effective time domain to a constant value of $\tau_0$. Therefore, the transformation of eq (2) from the real time domain to the effective time leads to BSP in the effective time domain given by:[45, 105]

$$\gamma(\xi) = \int_{-\infty}^{\xi} J(\xi - \xi_w)\frac{d\sigma}{d\xi_w}d\xi_w \text{ and } \sigma(\xi) = \int_{-\infty}^{\xi} G(\xi - \xi_w)\frac{d\gamma}{d\xi_w}d\xi_w, \qquad (4)$$

where $\xi_w = \xi(t_w)$ denotes the effective time at the application of deformation field. The existence of eq (4) also leads to the convolution relation in the effective time domain given by: $\xi = \int_0^\xi G(\zeta)J(\xi - \zeta)d\zeta$. Very interestingly Laponite dispersion has been observed to follow not just each of the expressions in eq (4) but also the convolution relation as evidenced from figure 7.[88] Furthermore, Laponite dispersion has also been observed to validate time-temperature superposition in the effective time domain.[107]

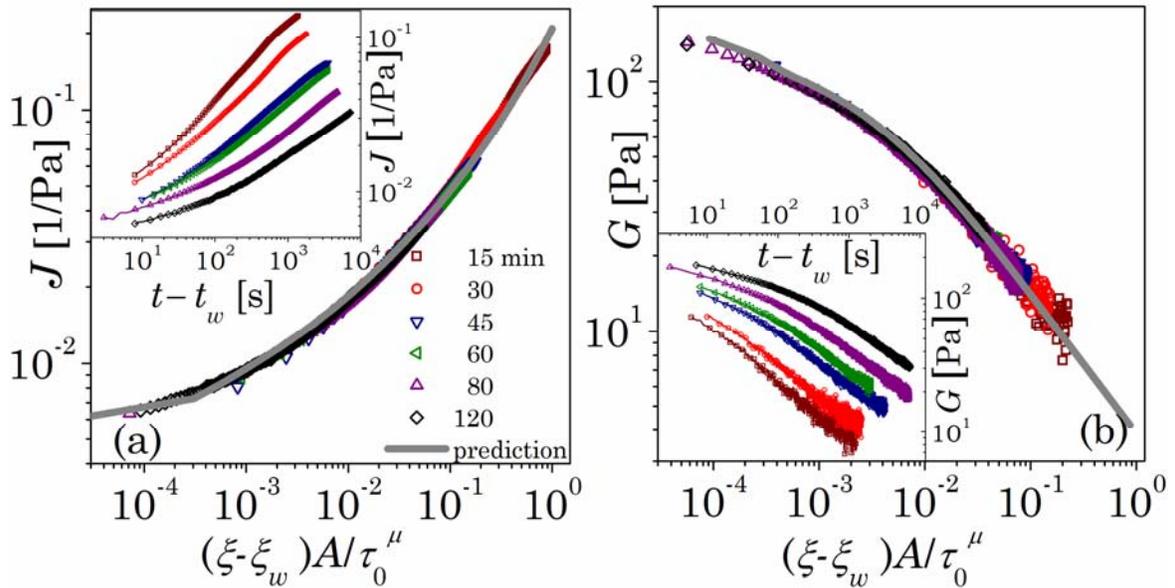

**Figure 7.** Superposed creep compliance (a) and stress relaxation modulus (b) curves corresponding to different $t_w$ in effective time domain for $C_L = 3.5$ wt. % with $\mu = 1.21 \pm 0.02$. The solid gray line in (a) denotes the predicted compliance from stress relaxation data while in (b) is the predicted stress relaxation modulus from compliance data using convolution equation in the effective time domain. The inset shows the evolution of compliance (a) and stress relaxation modulus (b) for the same system as a function of $t_w$. Reproduced with permission from reference 88. Copyright 2014 The Royal Society of Chemistry.

As shown in figure 7, validation of BSP also leads to an estimation of the dependence of relaxation time on the real time that requires carrying out creep or stress relaxation experiments at different $t_w$. It has been observed that the relaxation time of Laponite dispersion not just depends on $t_w$ but also on time over which the Laponite dispersion has been kept at rest before applying shear melting (rest time, $t_R$).[108] At small $t_R$, slow relaxation time shows an exponential dependence on $t_w$,[20, 108] however with the increase in rest time the dependence of relaxation time changes to a power law: $\tau_2 \sim \tau_m^{\mu-1} t_w^{\mu}$, where $\mu$ is power law exponent. Here $\tau_m$ is microscopic relaxation time that is necessary for dimensional consistency, which can be estimated experimentally.[96, 107, 109] It can be seen that this rheological behavior is qualitatively similar to the results of light scattering experiments. It has been observed that with an increase in $t_R$, power law exponent decreases from a value significantly greater than unity to unity over a rest time of the order of six months.[108]

The time dependent modulus and stronger than linear relaxation time dependence on time is an important characteristic feature of the aging behavior of Laponite dispersion. On one hand, stronger than linear relaxation time dependence prohibits complete relaxation of a material leading to residual stresses. On the other hand, in a creep flow field this feature leads to a plateau in the compliance. According to Joshi[110] stronger than linear relaxation time along with time dependent modulus leads to non-monotonic steady state shear stress – shear rate relationship. Interestingly such non-monotonicity is responsible for the observed time dependence of yield stress and yield strain as well as the shear banding, and model by Joshi[110] quantitatively predicts many behaviors shown by aqueous dispersion of Laponite. Many of the peculiar rheological behaviors shown by Laponite dispersion under non-linear flow field occurs because of non-trivial alteration of relaxation distribution. While many such rheological behaviors are qualitatively shown by Soft Glassy Rheology model,[84] quantitative description of the same has still not been possible. As mentioned before, many industrial systems employ Laponite as a rheology modifier that renders the former paste like consistency as well as thixotropy. Therefore, understanding soft glassy dynamics including rheology of Laponite dispersion is a key to understand the behavior of a spectrum of commercial as well as model time dependent systems.

**Microstructural Investigation:**

The previous section illustrates how aqueous dispersion of Laponite undergoes physical aging wherein its modulus and relaxation time continuously evolve with time eventually turning a free-flowing liquid into a soft solid. The microstructure that is responsible for such enhanced stiffness has been a subject of intense investigation over the past two decades. Various kinds of characterization techniques including scattering, microscopy, rheology, electrochemical analysis, etc., and simulations have been employed to study the same. In this section, we critically analyze key findings of such efforts.

1. Light Scattering and optical birefringence

Light scattering has been extensively used to investigate the phase behavior of aqueous Laponite dispersion. Light scattering is known to explore the length-scales over a range: 150 nm - 10 μm. Many groups employ static as well as dynamic light scattering to infer the microstructure. Avery and Ramsay[30] were the first to study aqueous Laponite dispersion using light scattering technique. They observe that translational diffusion coefficient of Laponite particles in low concentration regime (≈ 1 wt. %) remains practically constant with aging time whereas for high concentrations (≈ 3 wt. %), it decreases with time. By using DLS, Bonn et al.[51] investigate 3.5 wt. % Laponite dispersion and propose the system to be a repulsive (Wigner) glass. This was based on criteria mentioned by Bosse et al.[111] that suggests a possibility of formation of glass in a low density Coulomb system interacting via long range repulsive potential with an interparticle distance greater than the particle diameter. By taking into account the Debye screening length associated with the negatively charged faces of a particle an effective volume fraction of the particles has been estimated to be around $\phi \approx 0.5$ at which hard-sphere colloidal systems show a glassy state.[51] However, Mongondry et al.[3] have objected to the calculation of Debye screening length as it ignores the contributions from the counter-ions. Mongondry et al.[3] argue that effective volume fraction (that includes Debye screening length in the particle dimensions) of Laponite dispersion can never be near to close packing value for the concentration range of interest.

It has been observed that if Laponite dispersion is used without filtration, the static light scattering (SLS) experiments show a misleading power law decay in a limit of low $q$.[112, 113] Since filtration through submicron sized filters breaks the undissolved aggregates thereby avoiding false scattering signals, it has been customary to filter the Laponite dispersions. However, interestingly Bhatia et al.[114] observe a sharp increase in intensity with decrease in $q$ in Ultra small angle neutron scattering (USANS) experiments carried out on filtered 6 wt. % dispersions, which they attribute to significantly aged state of the same. Furthermore, the DLS experiments of Cummins[7] on 0.18 wt. % sample did not show any specific dependence of filter size on the rate of aggregation, which he attributes to the residual impurities already present in the filter. Moreover, Cummins[7] observe the initial dynamics to be slower in a sample filtered after 88 days than in a sample filtered immediately after the preparation suggesting the filtration process is unable to break all the clusters.[7]

One of the quantities of interest in scattering experiments is the behavior of intermediate scattering function $f(q,t,t_w)$ (with non-ergodicity parameter given by $f(q,t \to \infty,t_w)$), which has been regarded to give spatiotemporal information of the system. Jabbari-Farouji et al.[16, 48] investigate Laponite dispersions over a broad range of concentrations ($0.2 < C_L < 3.5$ wt. % and $0 < C_S < 7.5$ mM) using DLS and report $f(q,\infty,t_w)$ to reach unity for low concentrations ($C_L < 1.2$ wt. % and $1 < C_S < 7.5$ mM).

They attribute the same to rigidly held colloidal particles in place with very little freedom of movement suggestive of an attractive gel structure. On the other hand, at high concentrations (2.5 < $C_L$ < 3.5 wt. % and 0 < $C_S$ < 2 mM), they observe $f(q,\infty,t_w)$ to saturate at 0.8, suggestive of more freedom of movement, which they propose to be a repulsive glass. In the intermediate concentration range (1.2 < $C_L$ < 2.3 wt. % without salt) they claim equal possibility for the system to get arrested in either of the ways. Besides Jabbari-Farouji et al.[16, 115] suggests Laponite dispersion over the concentration range of 0.8 < $C_L$ < 1.5 wt. % with 5 < $C_S$ < 8 mM to be in the attractive glass state. Mongondry et al.[3] examine filtered samples over a range of Laponite (0.05< $C_L$ < 2 wt. %) and salt (0 < $C_S$ < 200 mM) concentrations, and based on their visual observations and SLS measurements they present a state diagram for the same. They propose Laponite particles to undergo aggregation composed of positive edge – to – negative face association. Importantly Mongondry et al.[3] report phase separation for $C_L$ <0.3 wt. %. However, Ruzicka et al.[116] report first gelation and then phase separation from the gel for $C_L$ <1 wt. %.

Ruzicka et al.[52] investigate aqueous Laponite dispersion over a broad concentration range (0.3 < $C_L$ < 3.1 wt. % and 0.1 < $C_S$ < 1 mM) using DLS. They fit the intensity autocorrelation function to eq (1) and observe that $\tau_2$ increases with $t_w$ and diverges at certain critical value $t_w^\infty$. Interestingly when $t_w$ is normalized by $t_w^\infty$, evolutions of $\tau_2$ as well as $\beta$ show two master-curves: one belonging to high concentration ($C_L$ > 1.8 wt. %) and the other belonging to low concentration ($C_L$ < 1.5 wt. %). They represent the intermediate region as a transition region. Based on this behavior Ruzicka et al.[52] propose that Laponite dispersion follows two different routes to reach an arrested state depending upon the concentration of Laponite. Cummins[7] inspect the effect of pH adjustment on relaxation dynamics of Laponite dispersion using DLS, and report that samples ($C_L$=1 wt. %) prepared without pH adjustment display faster aggregation compared to the samples whose pH was adjusted to 10 before addition of Laponite. As discussed before, lowering of the initial pH make the edges more electropositive. The DLS study by Cummins[7] therefore suggests that when edges are more electropositive, the aggregation rate is faster thereby endorsing the edge-to-face gel structure for the studied concentration of Laponite dispersion.

Interestingly for Laponite dispersion with $C_L \geq 2.8$ wt. % a permanent birefringence suggestive of nematic ordering has been reported in the literature.[6, 117] Very interestingly, Shahin et al.[117] observe development of optical birefringence at Laponite dispersion – air interface compared to bulk over a length-scale of several millimetres. The extent of ordering has been observed to be greater on any given day for a higher concentration of salt as well as at higher temperatures. Interestingly such enhanced ordering near the interface compared to the bulk was absent when free surface is covered with paraffin oil. Consequently, it is recommended that the free surface of Laponite dispersion in any experiment be covered with organic oil to avoid such an effect.

2. SAXS

Small angle X-Ray scattering (SAXS) has been routinely employed while characterizing the microstructure of soft materials. The advantage of small angle light scattering is that the range of $q$ it explores corresponds to length-scales from 1 nm to 100 nm that suits the typical interparticle length-scales observed in Laponite dispersions. In scattering studies, the structure factor ($S(q)$) is obtained by dividing intensity of scattered light ($I(q)$) by a form factor ($P(q)$. However, strictly speaking, representation of $I(q)$ as: $I(q) = n_p P(q) S(q)$, where $n_p$ is the number density, is only possible for suspension of monodispersed spherically shaped particles.[118, 119] However, for the anisotropic particles and/or particles with polydispersity, the decoupling of $I(q)$ into product of $P(q)$ and $S(q)$ is carried out assuming that system follows decoupling approximation and local monodisperse approximation.[118] In decoupling approximation, it is assumed that particles size, shape and orientation is independent of its position,[118, 120] while in local monodisperse approximation, it is presumed that over a short range all the particles are monodispersed. In addition to these assumptions, and by carrying out a variety of further approximations, SAXS has been routinely used to analyze colloidal dispersions having different kinds of microstructure by evaluating $S(q)$ as discussed in the classical reviews by Pedersen[121] and Li et al.[118]

More recently Greene et al.[122] critically analyze the validity of decoupling approximation for hard ellipsoids of revolution having aspect ratio greater than (prolate) and less than (oblate) unity. Greene et al.[122] infer that with an increase in aspect ratio and volume fraction of the particles, the error in obtaining structure factor through decoupling approximation increases significantly. Furthermore and very importantly, they conclude that at the higher concentrations and aspect ratios, the decoupling approximation leads to spurious peaks in the evaluated $S(q)$ that do not correspond to any characteristic length-scale in the system. They claim that for particle volume fraction beyond 1 %, the suggested validity of decoupling approximation is restricted to aspect ratio of 1/3. Laponite particles have a far greater aspect ratio of 1/30, which therefore raises very serious doubt over the analysis of SAXS of Laponite dispersions by evaluating $S(q)$. Moreover, if $2\alpha$ is the maximum length-scale associated with an anisotropic particle, Greene et al.[122] mention that below this length-scale the particles experience excluded volume interactions restricting certain rotations leading to coupling between their positions and orientations. Consequently, the system violates the decoupling approximation in that limit and the evaluated $S(q)$ cannot be used to decipher the microstructure. Therefore, for a Laponite particle decoupling approximation clearly gets violated for length-scales below 30 nm.

The analysis of Greene et al.[122] strictly refers to hard sphere interactions. The authors mention that when particles share attractive interactions, the relative positions and orientations of the same get more coupled, limiting the validity of the decoupling approximation even further. In an aqueous dispersion of Laponite having pH around 10, owing to dissimilar charges on edges and faces, as the two Laponite particles come close to each other they experience edge-to-face attraction as well as face-to-face repulsion. Consequently, there should be a significant correlation between the interparticle separation and their orientation particularly at high concentrations, where average interparticle length-scales are of the order of the diameter of a particle. As a result, the decoupling of $I(q)$ into the product: $P(q)S(q)$ for aqueous dispersion of Laponite may not be considered as the best approximation to *quantitatively* interpret peaks in $S(q)$ to determine the microstructure. We therefore believe that SAXS is an important tool for analysis for anisotropic systems with smaller aspect ratio. However, given that the strong approximations are involved in obtaining $S(q)$ for particles with high aspect ratio and charge anisotropy such as aqueous Laponite dispersion leading to possibility of spurious peaks and misleading results its qualitative as well as quantitative interpretation may not lead to reliable description of the microstructure.

Notwithstanding this discussion, assuming that accurate description of $S(q)$ is possible, Tanaka et al.[123] suggests some possible observations in $S(q)$ versus $q$ plot that may act as a signature to distinguish a repulsive glass from an attractive gel, particularly for aqueous dispersion of Laponite. According to Tanaka et al.,[123] a repulsive glass formed by Laponite particles is expected to show a peak at $q_p \approx 2\pi/\bar{l}$, where $\bar{l}$ is average interparticle distance. On the other hand, an attractive gel formed by Laponite particles is expected to show a peak between: $2\pi/d < q_p < 2\pi/h$, where $d$ is the diameter of a Laponite particle while $h$ is the thickness of a particle. For an attractive gel like structure, Tanaka et al.[123] further suggests that over a length-scale of $q\zeta << 1 << qd$, where $\zeta$ is the correlation length associated with the network, $S(q)$ is expected to show a power law tail: $S(q) \sim q^{-f_d}$, where $f_d$ is the fractal dimension associated with a gel. In addition, since osmotic compressibility is proportional to $S(q \to 0)$, Tanaka et al.[123] propose that respectively decrease and increase in $S(q \to 0)$ suggests an increase in repulsive interactions (leading to repulsive glass) and attractive interactions (leading to attractive gel).

SAXS technique has also been employed on Laponite dispersion by Levitz and coworkers,[124] who report the effective thickness of Laponite particle, as computed from SAXS data at high $q$ region where curves of different concentrations overlap, to be around 1 nm. Since the effective thickness is equivalent to the thickness of an individual particle, the authors conclude the absence of a house of cards structure. To examine the microstructure at very low ionic strength, Levitz and coworkers[125] study the evolution of effective structure factor using ultra SAXS technique, where they speculate a heterogeneous structure at a higher value of $C_L$.

Extensive investigation of aqueous dispersion of Laponite using SAXS has been carried out by Ruzicka et al.[126-128] They study low ($C_L$ < 2 wt. %) and high ($C_L$ > 2 wt. %) Laponite dispersion[126] without any externally added salt and reported the following observations that apparently matched well with the proposal of Tanaka et al.[123] For low concentration systems they observe a peak in $S(q)$ at $q$ associated with 15 nm, which suggests particles to form a space spanning percolated network having attractive interactions. However, for high concentration systems, the authors claim peak in $S(q)$ at $q$ associated with 40 nm, which they attribute to a repulsive glass-like structure.[126] However careful observation of the plotted $S(q)$ gives rise to an ambiguity about the sharpness of peak associated with both, low concentration as well as high concentration system.[126, 128] While the peaks associated with low concentrations dispersions at high $q$ is very weak,[126, 128] the peak at 40 nm is not prominent for 3 wt. % ( $t_w$ > 2000 min).[128] Furthermore, they report that in the low $q$ region, SAXS curve for low concentration dispersions shifts to higher values with increase in $t_w$[126] leading to power law tail as suggested by Tanaka et al.[123] However, such power law tail does not seem to be present in every low concentration Laponite dispersion studied by Ruzicka et al.[128] For high concentration systems, on the other hand, they observe a plateau of $S(q)$, whose value decreases with increase in $t_w$.[126] Ruzicka et al.[126] suggest decrease in scattering intensity with increase in $t_w$ to be a signature of repulsive glass state. However, Mongondry et al.[3] propose that such behavior is not incompatible with attractive gel state, wherein with an increase in $t_w$, there could be a gradual break-up of small dense Laponite clusters. Since the structure associated with "house of cards" is less dense than these clusters, the decrease in scattering intensity implies that Laponite clusters eventually break and then participate in "house of cards" structure.[3] Furthermore Ruzicka et al.[126] attribute the arrested state in high concentration sample to be glass owing to the presence of almost flat $S(q)$ in the low $q$ region. This $q$ independent structure factor behavior is interpreted in terms of a disconnected homogeneous structure at length-scales greater than the particle diameter, which is termed as Wigner glass. However, Mongondry et al.[3] argue that structure factor is not related to connectivity but instead it gives knowledge of the relative position of the particles at high concentrations. They further support their argument by giving an example of a crosslinked polymer that also shows $q$ independent $S(q)$ thereby arguing the observed behavior also to be compatible with an attractive gel structure. Overall the above discussion suggests that interpretation of SAXS results may either be misleading or ambiguous with respect to determination of definite microstructure for highly anisotropic and oppositely charged system such as Laponite dispersion.

3.  Microscopy

Various microscopic techniques have been employed over the years to investigate the microstructure of Laponite dispersion. Thompson and Butterworth[32] employ transmission electron microscopy (TEM) on 1 wt. % Laponite dispersion, wherein they observe aggregates of Laponite particles. However, owing to the difficulty in identifying the particles in the microscopic images, no conclusion on the phase behavior was drawn.[32] Mourchid et al.[15, 124] use cryo-TEM to explore microstructure of low (1 wt. %) and high (3 wt. %) concentration Laponite dispersions, which displays a homogeneous dispersion with well separated particles irrespective of the concentration.

Au and Leong[35] investigate 3 wt. % Laponite dispersion with 6 % pyrophosphate on dry weight basis using scanning electron microscopy (SEM) at pH 6 that demonstrated uniformly distributed large clusters, although the precise microstructure at an individual particle level is not apparent. Ranganathan and Bandyopadhyay[129] perform cryo-SEM on 3 wt. %, 0.5 mM NaCl and 3 wt. %, 0.5 mM $H_2SO_4$ dispersion, which displays a prominent dense network of Laponite particles. Subsequently, Gadige and Bandyopadhyay[130] employ cryo-SEM technique in 3 wt. % Laponite, where gelation is induced by the electric field, and observe overlapping coins and house of cards structures. The network connectivity is observed to increase with growth in electric field strength. In favor of these results, Jatav and Joshi[5] report cryo-TEM images of filtered low ($C_L$ = 1.8 wt. % with $C_S$ = 3 mM, 60 hours after preparation) and high ($C_L$ = 2.8 wt. %, 96 hours after preparation) concentration Laponite dispersion system. The cryo-TEM image of the low concentration sample studied by Jatav and Joshi[5] is shown in Figure 8. The images of both the systems indeed display interparticle edge-to-face bonds. Owing to the low concentration of the sample, a lean gel structure formed by single Laponite particles is observed as shown in Figure 8. However, the gel network is not explicitly visible as the space spanning network is expected to be present below the plane of observation. On the other hand, 2.8 wt. % dispersion indeed demonstrates a percolated structure in the plane of observation, which unequivocally establishes the system in both regimes to be in an attractive gel state. Interestingly, Jabbari-Farouji et al.[131] report high concentration (3.2 wt. %) dispersion to be homogeneous for length-scales greater than 0.5 μm. This observation matches very well with the cryo-TEM images of Jatav and Joshi,[5] who observe a homogeneous structure over a length-scale of 0.5 μm for 2.8 wt. % dispersion.

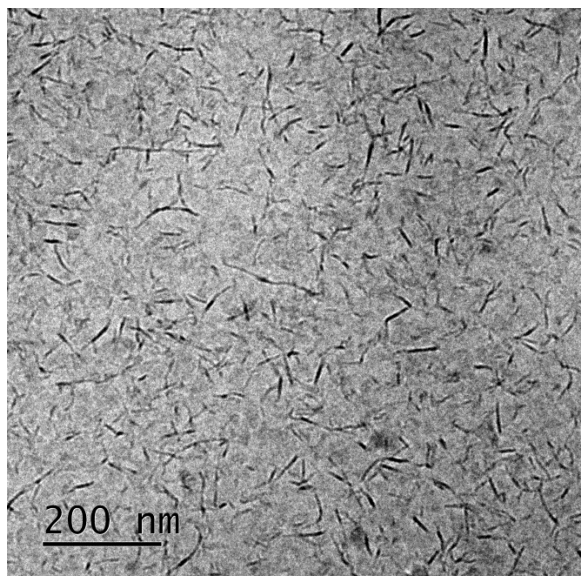

**Figure 8.** Cryo-TEM image of a mature gel of 1.8 wt. % Laponite with 3 mM NaCl at an age 60 hours. Reproduced with permission from reference 5. Copyright 2017 American Chemical Society.

In order to explore the microstructural evolution of high concentration Laponite dispersion as a function of time, we obtain the cryo-TEM image of 2.8 wt. %, no salt Laponite dispersion system 30 hours after preparation using the identical protocol and instrument as mentioned by Jatav and Joshi[5]. In figure 9 we show the cryo-TEM image of 2.8 wt. %, no salt Laponite dispersion after 30 hours (Figure 9(a, c, e)) and after 96 hours (Figure 9(b, d, f)) at different magnifications. The rheology experiments for $C_L =$ 2.8 wt. % dispersion, as seen in figure 10(c), suggest the rheological sol-gel transition for this system occurs at around 50 hours after preparation. It can be seen that while both the plots show very strong inter-particle interactions, 30 hours sample does not show a space spanning percolated network at least in the plane of view. The 96 hours sample, on the other hand, not just shows the space spanning percolated network but also shows particles to be more homogeneously distributed than that of 30 hours old dispersion. This suggests that as dispersion ages, particularly in the early period after preparation, homogenization of number density as well as network building occurs simultaneously. It is also important to note that the system never shows repulsion dominated microstructure such as repulsive glass over the explored times unlike what is claimed in various scattering and simulation papers.

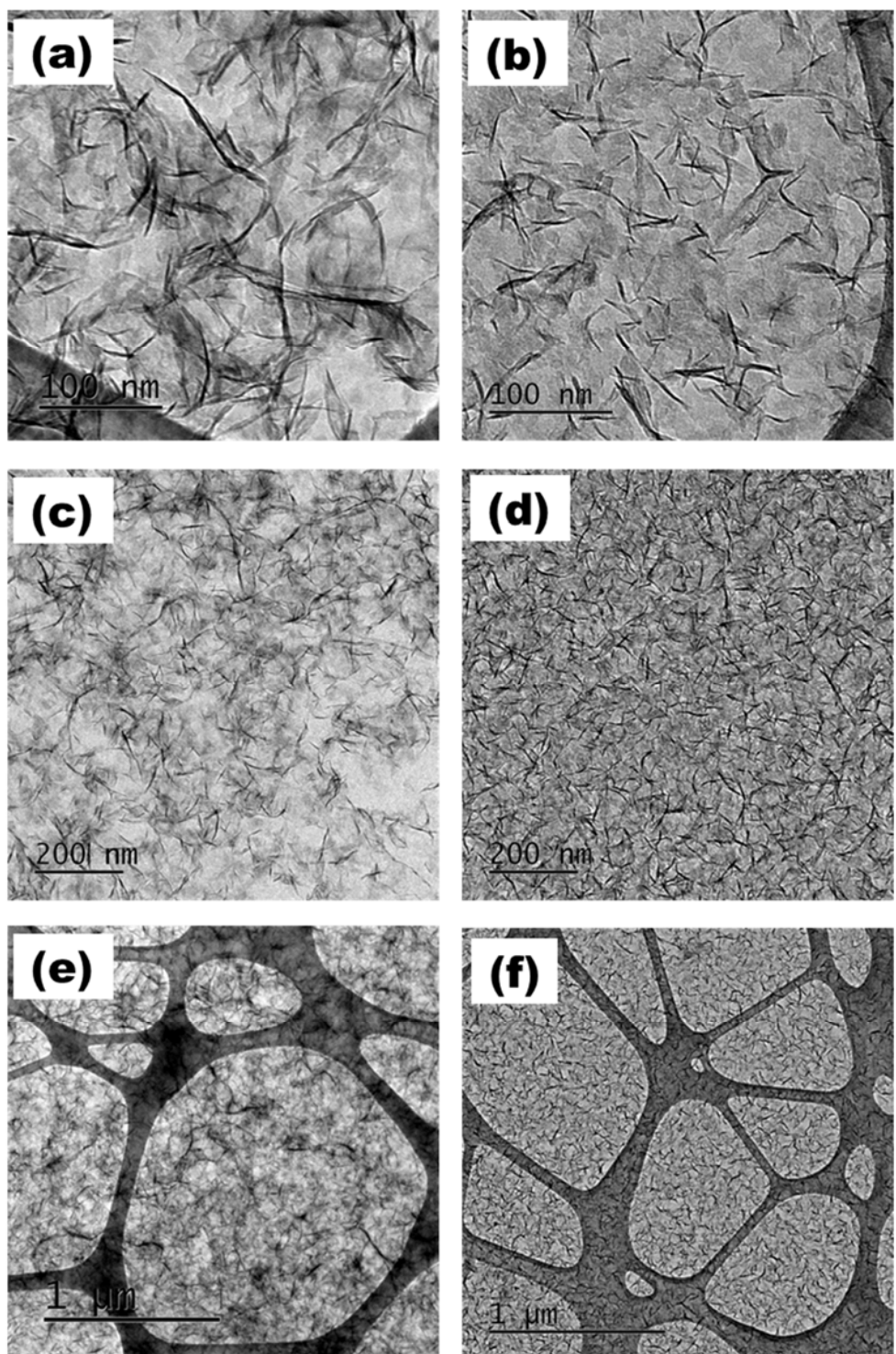

**Figure 9.** Cryo-TEM images of 2.8 wt. % Laponite dispersion at pregel state of age 30 hours (a, c, e) and postgel state of age 96 hours (b, d, f) at different length-scales.

4.  Rheology and Microrheology

Rheology is a mechanical tool, which apart from many other properties, measures the viscous as well as elastic character of a material for a given period of observation. In a sol-gel transition, since the material in a sol state (a liquid with dominant viscous response) converts into a gel state (a network like state with elastic properties), rheology gives an accurate description of this transition. It is known that in polymeric systems, monomers (sol state) undergo crosslinking reaction that eventually leads to the formation of a three-dimensional crosslinked network.[132, 133] In a seminal contribution Winter and Chambon (WC)[133] show that while in a sol (no network, liquid) – gel (dense network) transition, a system usually passes through the state with the weakest space spanning network having a fractal structure, which has been termed as a critical gel state. Winter and Chambon observe that at a critical gel state both $G'$ and $G''$ show a power law dependence on $\omega$ with identical exponent ($n$) given by:[133]

$$G' = G'' \cot\left(n\pi/2\right) = \frac{\pi S}{2\Gamma(n)\sin\left(n\pi/2\right)} \omega^n, \tag{5}$$

where $S$ is the gel strength, $\Gamma(n)$ is the Euler gamma function of critical exponent $n$ having limits: $0 \leq n \leq 1$. Consequently, at a critical point loss tangent $\left(\tan\delta = \tan(n\pi/2)\right)$ is independent of $\omega$. In addition to numerous polymeric systems,[132-137] WC criterion has also been observed to be followed by the colloidal systems that undergo sol – gel transition[5, 138]. Moreover, the WC criterion has not just been confirmed for bulk rheological techniques (probed length-scale > 500 μm)[5, 8] but also for microrheological techniques (probed length-scale < 2 μm)[73-75, 131] suggesting it to be a robust criterion to establish a sol – gel transition. For a three-dimensional gel, Muthukumar[139] has related $n$ to the fractal dimension $\left(f_d\right)$ of a critical gel by: $f_d = 5(2n-3)/2(n-3)$. Very remarkably, this observation has been validated for many systems by carrying out simultaneous rheology as well as scattering experiments.[140-142] A primary thesis of observation is that being a 'real' transition associated with the network formation, the point of gelation transition is independent of the timescale of observation that leads to $\tan\delta$ being independent of $\omega$ (frequency is inverse of the timescale of observation). On the other hand, the glass transition, which also describes liquid – solid transition, strongly depends on the timescale of observation.[90]

Furthermore, at the point of sol–gel transition, since the space spanning network is the weakest, the longest relaxation mode is least populated while the faster relaxation modes gradually dominate. As a result, the relaxation time spectrum shows an inverse power law dependence.[143] This is in contrast to the glass transition, where the slower modes are densely populated. Therefore, on one hand, rheology very clearly distinguishes the sol–gel transition from the glass transition. On the other hand, it predicts the microstructural attributes of a critical gel state suggesting rheology to be a vital tool to characterize the microstructure of soft materials, particularly the liquid–soft solid transition in Laponite dispersion.

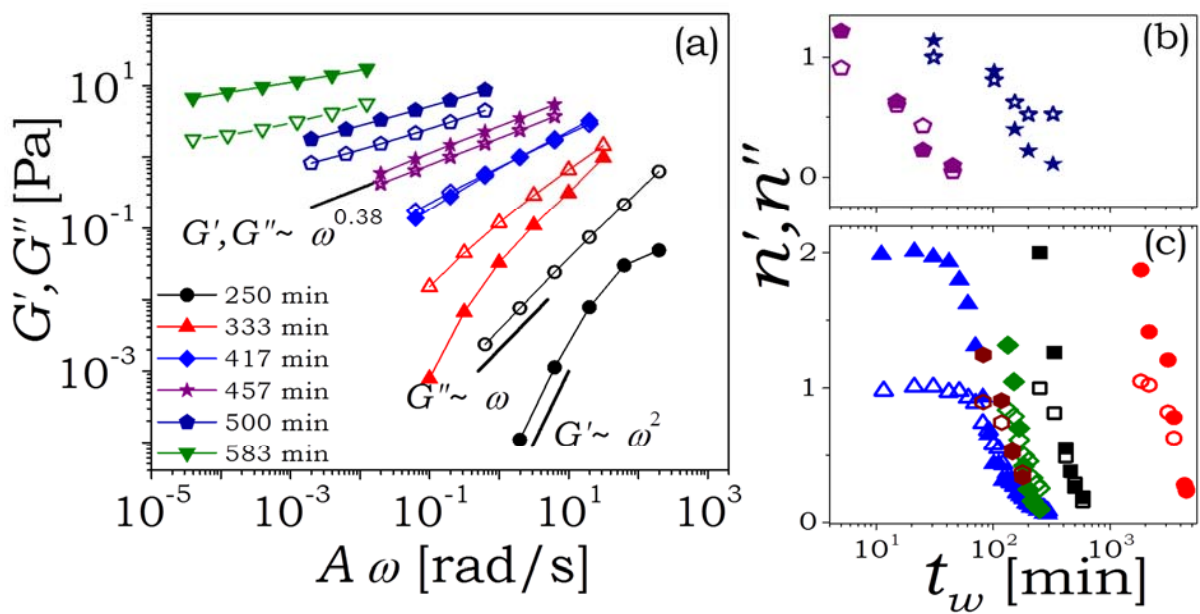

**Figure 10.** (a) Temporal evolution of $G'$ and $G''$ with angular frequency $(\omega)$ at different aging times for 2.8 wt. % Laponite dispersion with 3 mM NaCl. For clarity, the curves have been shifted along the horizontal axis by varying parameter $A$. Reproduced with permission from reference 8. Copyright 2014 The Society of Rheology. Evolution of power law dependence of the viscoelastic moduli on $\omega$ $\left(G' \sim \omega^{n'}, G'' \sim \omega^{n''}\right)$ with aging time for the data from different groups for a broad range of $C_L$ and $C_S$ using (b) the microrheological measurements and (c) the bulk measurements. The legend for (b) and (c) are mentioned in Table 1 and 2.

For the first time validity of the WC criterion for freshly prepared low concentration Laponite dispersion was confirmed by Nicolai and coworkers[138] ($C_L$=1 wt. % and $C_S$=5 mM) while for high concentration Laponite dispersion the same was confirmed by Coussot and coworkers[43] ($C_L$=3.5 wt. %). Table 1 summarizes all the systems explored in the literature, which validate the WC criterion[133] using bulk rheology. In figure 10(a) we plot $G'$ and $G''$ as a function of $\omega$ at different times since preparation of the dispersion of $C_L$=2.8 wt. % and $C_S$=3 mM. In figure 10(c) we plot the evolution of $n'$ and $n''$ ( $n' = d\log G'/d\log \omega$ and $n'' = d\log G''/d\log \omega$ ) as a function of time for the bulk measurements data from different groups for a broad range of $C_L$ and $C_S$. It can be seen that the response corresponding to typically that of a sol ( $G' \sim \omega^2$ and $G'' \sim \omega$ ) changes with time in such as fashion that at a certain point both $G'$ and $G''$ show identical dependence on $\omega$ given by eq (5). Very remarkably, irrespective of a group, geometry employed as well as the length-scales probed, $n'$ and $n''$ indeed shows a similar trend. Since modulus of a material is represented as energy density, increase in instantaneous modulus ($G'$ in a limit of high $\omega$) as a function of time suggests an increase in bond energy as well as densification of a network as a function of time. In addition, Jatav and Joshi[5] systematically examine the freshly prepared aqueous dispersion over a broad range of $C_L$ (1.4 to 4 wt. %) and $C_S$ (0 to 7 mM NaCl), and reported validation of the WC criterion for the entire domain. The associated critical gelation time $\left(t_g\right)$ for the wide range of systems studied by Jatav and Joshi[5] has been shown in figure 11(a).

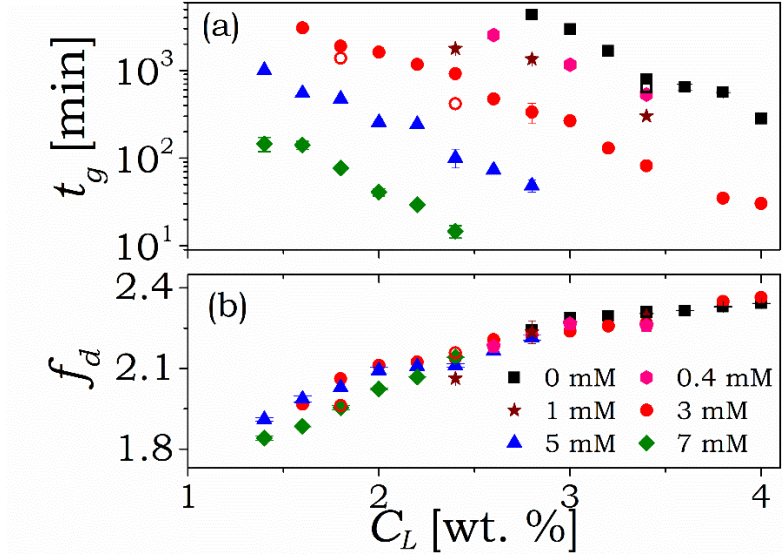

**Figure 11.** (a) Evolution of critical gelation time $(t_g)$ with Laponite concentration $(C_L)$ at different concentrations of salt $(C_s)$. (b) Evolution of fractal dimension $(f_d)$ associated with the critical gel state as a function of $C_L$ for different $C_S$ (0-7 mM). Open symbols represent the filtered dispersion. Data from reference 5.

The rheological features of a critical gel state provide an important information about the nature and kinetics of the same. Jatav and Joshi[5] report that the time $(t_g)$ at which a critical gel state is observed decreases exponentially with $C_L$ as well as $C_S$. In the polymer literature, the gelation is considered as an activation process, and the corresponding activation energy is obtained by assuming Arrhenius dependence of $t_g$ on the same:[144, 145] $t_g = t_{g0} \exp(E/k_B T)$ with prefactor representing time to gelation in a limit of negligible activation energy or very high temperature. Jatav and Joshi[5] suggest that the activation energy to form a gel $(E)$ decreases linearly with $C_L$ as well as $C_S$ according to: $E \sim -k_B T [2.14 C_L + 0.84 C_S]$, with $C_L$ in wt. % and $C_S$ in mM. They also observe $t_g$ to decrease with temperature as was observed for the polymeric systems. At ambient temperature, the activation energy for polymeric gels has been reported to be in a range 40 to 100 kJ/mol,[137, 146] while for Laponite dispersion (2.8 wt. % with 3 mM salt) the same is observed to be 245 kJ/mol.[5] The critical exponent $n$ also leads to the fractal dimension of the critical gel state. In figure 11(b) we plot $f_d$ as a function of $C_L$ having different $C_S$. It can be seen that $f_d$, which is observed to be in a range of 1.8 to 2.4, increases with increase in $C_L$ as expected, but shows a weak dependence on $C_S$. The fact that $t_g$ is a strong function of $C_S$, weak dependence of $f_d$ on the same suggests that the presence of salt affects the kinetics rather than the structure of Laponite dispersion.[5, 22]

The gel strength $S$ in eq (5) has dimensions of Pas$^n$. The literature on polymeric gels suggests: $S = G_0 \tau_0^n$, where $G_0$ has been proposed to be modulus associated with the densest critical gel while $\tau_0$ has been proposed to be the relaxation time of a polymer precursor unit.[147] Very interestingly, all the $\ln S$ versus $n$ data for the Laponite dispersions over a broad range of $C_L$ and $C_S$ has been observed to collapse on a single curve.[5] Moreover slope of $\ln S$ versus $n$ for $C_L$ below and above 2 wt. % has been observed to be distinctly different suggesting different kinds of gel structures in these two domains. Furthermore, estimation of $\tau_0$ in a limit of small $C_L$ leads to a timescale associated with the inverse of rotational diffusivity associated with a single Laponite particle. This observation suggests that in a low concentration system ($C_L < 2$), a single Laponite particle forms a precursor unit in the network. Jatav and Joshi[5] also obtain various critical exponents, whose values were also observed to match very closely to that of obtained for a very broad range of polymeric systems suggesting in every possible way, the process of gelation of in Laponite dispersion is akin to that observed for polymeric gel systems.

Interestingly, the critical sol–gel transition has been observed for a Laponite dispersion only when a system undergoes spontaneous evolution.[8] If a spontaneously evolved Laponite dispersion deep into a matured gel state is subjected to strong deformation field (rejuvenation), it undergoes yielding leading to a liquid state. Upon removing the deformation field, it again evolves and undergoes solidification. However, during this solidification process system is not observed to show a critical gel state. Moreover, the evolution of $G'$ and $G''$ subsequent to shear melting carried out at a later date since preparation of dispersion has been observed to shift to smaller times. This behavior has been attributed to the inability of shear melting to completely break the inter-particles bonds. Very interestingly, subsequent to shear melting of a matured gel, the relaxation time distribution has been observed to show a positive slope that is known to be a characteristic feature of a glass state. This observation has been attributed to the formation of undeformed solid pockets sliding past each other during the rejuvenation.[8] However, even in this case, the building block of the microstructure has still been proposed to be the interparticle bonds between the Laponite particles. The physical reasoning behind these observations has been discussed in further details in the DLVO section below.

In addition to bulk rheology, micro-rheology has also been used by various groups to assess aqueous dispersion of Laponite. It has been widely observed that the rheological behavior (or the mean squared displacement of the probe particles) of aqueous Laponite dispersion strongly depends on the length-scale of a probe particle.[73, 74, 131, 148, 149] Very importantly, similar to that observed for the bulk rheological techniques, many groups have reported validation of WC criterion[133] using microrheology for high ($C_L >$ 2 wt. %)[75, 131] and low concentration ($C_S <$ 2 wt. %)[73, 74, 131] over the length-scale of 1 to 3 μm as listed in Table 2. Figure 10(b) illustrates the evolution of $n'$ and $n''$ as a function of time for the dataset obtained by the microrheological measurements. It can be seen that there is a striking similarity between the bulk behavior and microrheological behavior over a broad range of probe particle length-scales suggesting universality of the WC criterion in probing the sol–gel transition.

Table 1: List of systems which validate Winter-Chambon criteria through bulk Rheology

| Serial Number | Laponite concentration (wt. %) + Salt concentration (mM) | Geometry | Geometry Gap (mm) | Reference | Legend in figure 10 (c) |
|---|---|---|---|---|---|
| 1. | 1+5 | Cone and plate (1°) |  | Nicolai and coworkers[138] | Green squares |
| 2. | 3+0 | Couette | 1 | Coussot and coworkers[43] | Blue up triangle |
| 3. | 3+0 | Couette | Not mentioned | Strachan et al.[148] | Brown hexagons |

| 4. | 2.8+3 | Couette | 1 | Jatav and Joshi[5, 8, 22] | Black squares |
| 5. | (1.4 to 4)+(0 to 7) | Couette | 1 | Jatav and Joshi[5] | 2.8+0: Red circle |

Table 2: List of systems which validate Winter-Chambon criteria through microrheological measurements

| Serial Number | Laponite concentration (wt. %) + Salt Concentration (mM) | Probe particle length-scale (μm) | Reference | Legend in figure 10 (b) |
|---|---|---|---|---|
| 1. | 0.8+6, 3.2+0 | 0.5, 1.16 | Jabbari-Farouji et al.[131] | 3.2 wt% : Navy blue star |
| 2. | 1+1 | 0.5 | Oppong et al.[73] | |
| 3. | 1+5.9 | 0.926 | Rich et al.[74] | Purple pentagons |
| 4. | 3+0.1 | 0.21 | Pilavtepe et al.[75] | |

5.  Dilution studies

The microstructure of aqueous Laponite dispersion, whether a repulsive glass or an attractive gel has been studied by dissolution experiments wherein Laponite dispersion in the semisolid state was kept in contact with deionized water. Mongondry et al.[3] conduct dissolution experiments on aged 2 wt. % Laponite dispersion and analyze the same using light scattering. They observe intensity to decay rapidly when dilution is carried out after 2 days. However, the decrease in intensity was slow for the diluted aged samples. They claim that this scenario could not be explained if interparticle interactions are repulsive in nature, but the slow breakage of aggregates may have led to the same.[3] Ruzicka et al.[127] carry out dilution experiments on 1.5 and 3 wt. % Laponite dispersions. The dilution experiments on 1.5 wt. % dispersion that they conduct after ergodicity breaking (several weeks after sample preparation) did not cause any change to the semisolid state of the same. Furthermore, the dilution experiments on the 3 wt. % dispersion after 50 hours of preparation (time associated with ergodicity breaking) results in melting of the same, while dilution experiment performed one week after the ergodicity breaking leads to swelling (penetration of water from the bulk) of the same. In a subsequent work, Ruzicka and coworkers[150] have performed dissolution experiments on 3 wt. % dispersion, 1, 2, 3 and 4 days after ergodicity breaking. They observe that while dissolution experiment did lead to melting of Laponite dispersion on day 1 and 2, the sample was unaffected on day 3 and 4. The authors claim that the low concentration dispersion remains unaffected in the dissolution experiments because of the formation of an attractive gel state due to interconnected clay bonds. They further claim, since the high concentration system is in a repulsive glass state on day 1 and 2, dissolution experiments leads to melting of the same. On day 3 and 4, however, the presence of disconnected house of card structure is claimed, which owing to prevailing attraction does not allow any effect on the semisolid state. Contrary to the above argument made in favor of high concentration system being in repulsive glass state, Mongondry et al.[3] argue that the interparticle positive edge–negative face bonds as well as the attractive van der Waals interactions that lead to the network are not irreversible. They further contend that the very fact that Laponite gels can be easily broken on the application of small stress suggest, the interparticle bonds are weak. As a result, upon dilution additional water can break the percolated gel structure and cause melting. Therefore, careful study of the literature suggests that dilution studies do not conclusively confirm the particular microstructure as observations can be interpreted either way.

6. DLVO and Conductivity measurements

The interaction energy between two semi-infinite plates approaching in a parallel fashion is given by the classical DLVO theory in terms of the normal distance between them.[2] According to this theory, total free energy of interaction per unit area $(W)$ is given by: $W = W_{DL} + W_{vdW}$ where $W_{DL}$ is contribution from the double-layer repulsion and $W_{vdW}$ corresponds to the van der Waals attraction. For computing $W_{DL}$, the surface charge density $(\psi)$ and the Debye screening length $(1/\kappa)$ are required while $W_{vdW}$ can be calculated by knowledge of Hamaker constant of the system. For a dispersion of Laponite in aqueous media, $\psi$ and $1/\kappa$ can be estimated by measuring pH and conductivity of a Laponite dispersion. In the inset of figure 12, we plot $W_{DL}$, $W_{vdW}$ and $W$ as a function of $\kappa d$, where $d$ is half distance between two plates.[10] It can be seen that as two plates approach each other in a parallel fashion (decrease in $\kappa d$), electrostatic interactions cause increase in repulsion while van der Waals interactions cause an increase in attraction. The intensity of $W_{vdW}$ can be seen to be significantly weaker than $W_{DL}$ except when both the particles are very close to each other. Consequently, the sum of both the interactions shows a maximum suggesting the dominance of repulsion when their faces are exposed to each other. The Debye screening length for the data shown in the inset of figure 12 is 3.8 nm, which suggests that the total interaction energy is predominantly repulsive at separation distance above 1 nm when particles approach each other in a parallel fashion. However the repulsive potential decreases with $d$ and becomes very small for $\kappa d \geq 1$. Figure 12 shows a schematic representation of $W$ as a function of $d$, wherein the height of the repulsive barrier (proportional to $\kappa$) can be seen to be increasing while its width (proportional to $1/\kappa$) can be seen to be decreasing with increase in the concentration of cations in the dispersion.[10] Concentration of cations may increase because of following reasons: increase in concentration of Laponite ($C_L$, that causes increase in Na$^+$ counter-ions), concentration of salt ($C_S$, addition of any salt with monovalent or divalent cation/anion combination), time (dissociation of Na$^+$ ions from the surface of the particles continues to increase ionic strength of the dispersion) and/or temperature (increase in temperature causes faster dissociation of Na$^+$ counter-ions).[10, 22, 42]

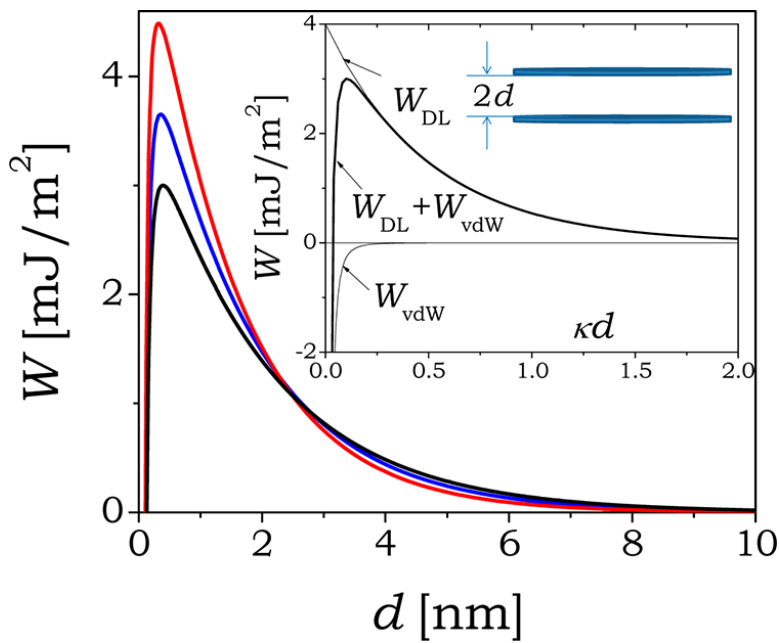

**Figure 12.** Variation of free energy per unit area of interaction between two parallel facing Laponite particles as a function of the distance between the particles at different times in 3.5 wt. % Laponite dispersion. The black, blue and red lines correspond to Day 3 ($\kappa^{-1} = 3.8$ nm), 12 ($\kappa^{-1} = 3.3$ nm) and 18 ($\kappa^{-1} = 2.9$ nm) respectively. The inset shows the contribution of van der Waals interaction and double layer repulsion to the total free energy per unit area as a function of normalized distance $\kappa d$ for 3.5 wt. % Laponite dispersion on Day 3. Reproduced with permission from reference 10. Copyright 2012 American Chemical Society.

The case discussed in figure 12 corresponds to a limiting case of two particles approaching each other in a parallel fashion. However parallel approach is just one of the possible scenarios, as in reality Laponite particles can approach each other in any orientation. Owing to the thermal motion exhibited by the particles, the most probable scenario of approach of any two particles is a non-parallel. Even for the parallel configuration, the length-scale over which the particles experience repulsion decreases with an increase in $C_L$, $C_S$, time, and $T$. Therefore it is not surprising that a non-parallel approach, along with the enhanced electronegativity of the faces with increase in $C_L$, $C_S$, time, and $T$ will facilitate faster formation of the positive edge–to–negative face bonds.[5, 22] This qualitatively explains the growth of $G'$ and $G''$ in Laponite dispersion due to faster gel formation as well as enhancement in slow and fast relaxation times with increase in $C_L$, $C_S$, time, and $T$ as discussed in rheology and light scattering sections and as shown in figures 5 and 11(a).[10] The observations of the DLVO analysis and its implications in faster structure formation when there is decrease in $1/\kappa$ is in intuitive disagreement with the proposal of repulsion dominated structure. If the structure would have been repulsion dominated its formation would have been slower with decrease in repulsive interactions. The experimental observations, therefore, clearly suggest that the structure must be inherently attractive in nature.

The DLVO analysis assumes that upon dispersing Laponite in water, delamination of the Laponite particles from tactoids is fast and dissociation of $Na^+$ counter-ions is a prolonged process. However, the above discussion also applies to the case when dissociation of $Na^+$ counter-ions is faster than delamination of Laponite particles. In the later scenario, the concentration of $Na^+$ counter-ions will gradually increase with progressive delamination and will show the same effect as shown in figure 12. However, in such case, the average interparticle distance of between the particles will also decrease as increasingly more particles will undergo delamination leading to an additional contribution to elastic modulus. While DLVO analysis can, in principle, qualitatively predict the interactions between charged particles in the presence of monovalent salt,[78] the analysis fails when multivalent ions are present in the system.[151] This is because the mean-field approximation of Poisson-Boltzmann equation fails to capture the additional interactions such as the ion-ion correlations, ionic dispersion forces, specific ion effects that arise in the presence of multivalent salts.[151, 152] However efforts have been made in the literature to account for the above effects in simulations.[151]

This section suggests that more and more Laponite particles form edge to face bonds as a function of time. The simulations as well as cryo-TEM images have indicated that these bonds could be in the edge to face configuration (not necessarily in the perpendicular orientation) as well as in the PPO (discussed below). As time passes, owing to thermal motion, the particles adjust their orientations and positions in a fashion to strengthen the bonds (that is akin to lowering of free energy with time). In addition, with time, more bond formation also takes place thereby increasing the bond density. Under such circumstances if Laponite dispersion is subjected to deformation field, only the fraction of the bonds formed during this evolution break leading to the formation of undeformed solid pockets that simply rub past each other. The size of such domain increase with time elapsed since preparation of Laponite dispersion. As a result, if Laponite dispersion is deformed on a progressively later date since its preparation, the fraction of the dispersion that cannot be broken by shear goes on increasing making the process of electrostatic bond formation between the particles irreversible. This property has been termed as irreversible aging in the soft glassy rheology literature.[8, 87] Furthermore, increase in $T$ enhances the thermal motion of the Laponite particles and small tactoids. Consequently, inter-particle bond formation gets accelerated at higher $T$. As a result, the evolution of $G'$ and $G''$ shift to lower times for experiments carried out at higher $T$ as shown in figure 5.[10]

7. Effect of multivalent salts and hydrophilic polymers

Introduction of salt in Laponite dispersion is known to profoundly influence the aging dynamics. As a result, studies exploring the nature of salt - clay interactions in the aqueous media have received significant importance. Out of various salts, most of the studies on Laponite dispersion investigate the effect of NaCl on the same.[5, 12, 97, 113, 123, 153] As a result, most of the discussion in this article involves the effect of NaCl on the microstructure as well as aging behavior of Laponite dispersion. Other than NaCl, the impact of other monovalent, as well as multivalent salts on Laponite dispersion, has been studied in the literature.[154] It has been observed that incorporation of most of the salts (except those containing anions with valency greater than 3) in Laponite dispersion leads to acceleration of the aging dynamics. As discussed before, various groups suggested that in the process of aggregate formation eventually leading to space spanning network, the presence of salt reduces the repulsive energy barrier needed for particles to approach each other resulting in faster kinetics. To best of our knowledge, there is only one report on the effect of multivalent salt on aging dynamics of Laponite dispersion. Thuresson et al.[154] investigated the effect of $CaCl_2$, $MgCl_2$ and $LaCl_3$ along with NaCl on aggregation of Laponite particles by SAXS, cryo-TEM and molecular dynamic simulations. The SAXS spectra for multivalent salts almost superimposed into a single curve thus concluding that all cations with valency >1 had a similar effect on the microstructure.

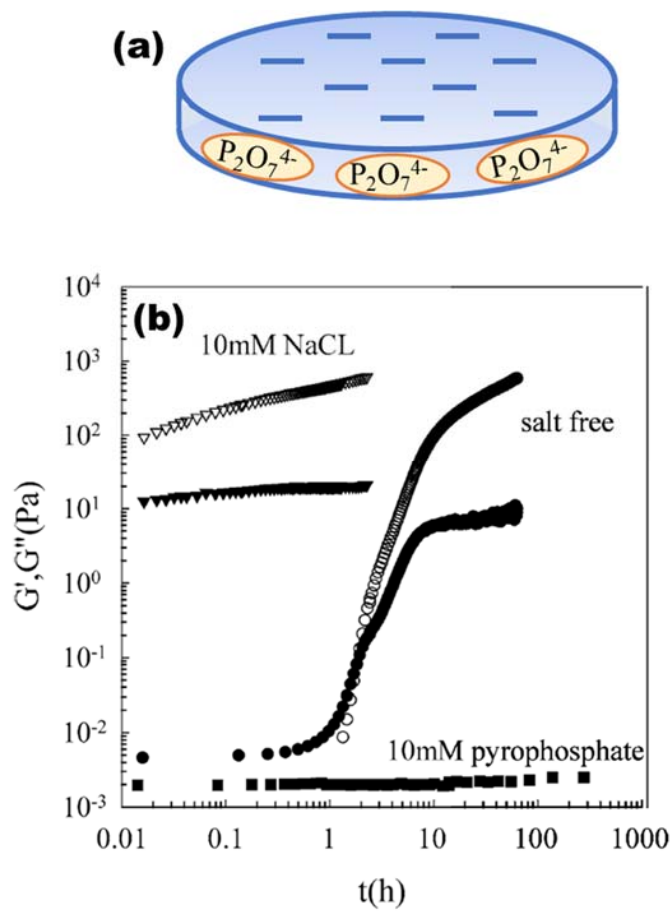

**Figure 13.** Schematic representation of the effect of (a) Tetrasodium pyrophosphate on Laponite particle in an aqueous medium. (b) Temporal behavior of $G'$ (open symbols) and $G''$ (filled symbols) in 2.5 wt. % Laponite dispersion with 10 mM NaCl, 10 mM pyrophosphate and salt free conditions. Figure 13(b) reproduced with permission from reference 155. Copyright 2004 Elsevier.

Interestingly the salts that contain anions with a valency greater than 3 such as tetrasodium pyrophosphate, sodium tripolyphosphate, sodium tetraphosphate and sodium hexametaphosphate have been observed to retard the process of aggregation.[156] Mongondry et al.[155] studied the influence of tetrasodium pyrophosphate on aggregation of Laponite dispersions using rheology and light scattering. Similar to the other salts, dissociated tetrasodium pyrophosphate does reduce electrostatic repulsion among the particles. However, more interestingly, the tetravalent anion $(P_2O_7)^{-4}$ gets adsorbed on the edge of Laponite particle thereby reducing the net positive edge charge as shown in figure 13(a). This slows down the rate of aggregation in Laponite dispersion. Effect of pyrophosphate on aging dynamics is clearly illustrated in Figure 13(b) which shows the evolution of $G'$ and $G''$ with time for three conditions: 2.5 wt. % Laponite dispersion without salt, 10 mM NaCl and 10 mM pyrophosphate. While 10 mM NaCl system undergoes gelation sooner than 0.01 hours, no salt system underwent gelation at about 2 hours. No gelation has been observed for 10 mM pyrophosphate system. The scattered intensity plots by Mongondry et al.[155] suggest a decrease in aggregation rate with increase in pyrophosphate concentration from 0.04 mM to 0.4 mM. Such inhibition in aggregation in the presence of pyrophosphate is also observed by Martin et al.[37] using light scattering and Au and Leong[35] using rheological tools.

Addition of hydrophilic polymer to aqueous Laponite dispersion is also known to alter the interparticle interactions. Effect of addition of polyethylene oxide (PEO) in Laponite dispersion has been first investigated by Lal and Auvray[157] using SANS. The authors speculate that either the polymer prevents gelation in 2 wt. % or slows down the aggregation process. Nelson and Cosgrove[158] introduced a core-shell model to account for a Laponite particle wrapped with the polymer. Later Mongondry et al.[155] incorporated SLS and DLS to examine the effect of PEO on the aging of 2 wt. % Laponite dispersion. The authors indeed observe an inhibition in the aggregation rate of Laponite dispersion. They suggest that the PEO chains adsorb on the Laponite particle that causes a reduction in the edge–face interaction. However, beyond a certain molar mass of PEO, Mongondry et al.[155] propose bridging between Laponite particles, which leads to the formation of clusters. Such bridged Laponite-PEO structure has also been reported by Thuresson et al.[154], Loizou et al.[159] and Baghdadi et al.[160] The delay in formation of the arrested state in Laponite-PEO dispersion has been monitored using rheology,[153, 160] DLS[60, 161, 162] and SANS studies.[155, 157, 158]

Atmuri et al.[163] studied the effect of molecular weight and concentration of PEO on aging behavior of Laponite dispersion. They propose that 2 wt. % Laponite dispersion, which they argue to be a repulsive glass in the presence of low concentration of PEO, transforms to an attractive glass with an increase in PEO concentration. However, with an increase in molecular weight, the authors report the formation of a gel. In a subsequent publication, Atmuri and Bhatia[161] study the effect of non-absorbing anionic polymer, polyacrylic acid, that results in slowing down of aging dynamics in Laponite dispersion.

8.   Simulations

Simulations are an important tool to investigate the microstructure and significant work has been carried out in the literature to understand the microstructure of Laponite dispersion. Simultaneous investigations using experiments and simulations on one hand help to decipher the physical mechanisms behind the observations. On the other hand, comparison with experiments leads to an accurate determination of simulated model characteristics including the interparticle potentials. Particularly for the aqueous dispersion of Laponite, owing to the anisotropic shape, dissimilar charges on the edges and faces, and profound effect of the nature of aqueous media, carrying out the realistic simulations has been a challenging task. Dijkstra et al.[164, 165] initiated the modelling study of synthetic clays. With the aim of proposing a quantitative statistical description, the authors consider an isolated Laponite particle in suspension as an infinitely thin hard disk carrying a rigid point quadrupole at their center. This quadrupolar interaction, which captures the orientation of neighboring platelets, was absent in the previously reported Monte Carlo (MC) simulations work of Eppenga and Frenkel which reported isotropic to nematic phase transition at 32 wt. %.[166] In order to avoid electrostatic collapse, Dijkstra et al.[164] introduce an additional infinite barrier term for center-to-center distances smaller than the radius of a disk. This repulsive barrier did not affect the T-shaped configurations, which was responsible for the gel network. The Quadrupole disk model gives rise to house-of-cards structure as the T-shaped configurations had the lowest energy and were energetically more favorable. In a subsequent work by Dijkstra et al.,[165] sol-gel transition of charged Laponite particles has been studied over a wide range of clay concentration by extensive MC simulations. The results indicate the sol-gel transition to be reversible first-order phase transition and yields the critical point above which sol-gel transition takes place to be 6 wt. %,[165] which is clearly an overestimation compared to the experimental results. However in Quadrupole disk Model, quadrupolar moment affiliated to the electric double layer is assumed to be independent of concentration.[167] In reality, at very short range double layers of adjoining platelets would overlap and the concept of point quadrupole would collapse.[167] On the other hand at long range, the model fails to explain the electrostatic interactions due to screening effects.[167] Furthermore neglecting the ubiquitous van der Waals interactions could lead to qualitative and quantitative changes in phase behavior of Laponite dispersion.[165] The Quadrupole disk Model, although rudimentary, is capable in predicting the sol-gel transition. Slightly more comprehensive models have been investigated in the literature where a Laponite particle is modelled as a pseudo site with appropriate size.

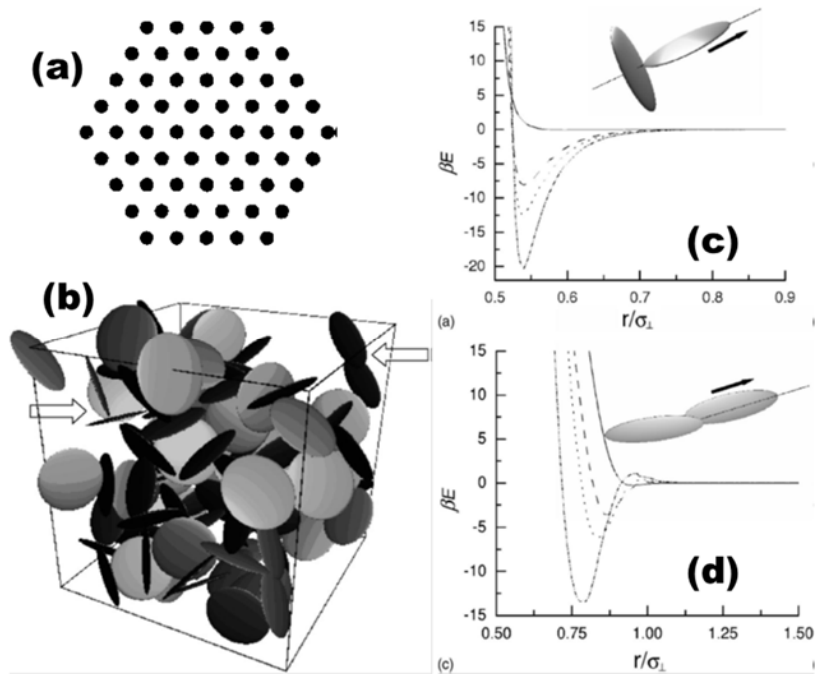

**Figure 14.** (a) Schematic representation of the model used by Kutter et al.[167] to represent one Laponite particle of diameter $D$. The model consists of 61 charged sites with $D/8$ spacing between sites. Reproduced with permission from reference 167. Copyright 2000 AIP Publishing. (b) Snapshot of configuration obtained by Odriozola et al.[168] for model B (accounts for negative face charge and positive edge charge) in 12.6 wt. % and $\kappa^{-1}$= 1 nm. The left and right arrow point towards a T-shaped and PPO configuration respectively. Variation of potential energy for two Laponite particles in model B with $\kappa^{-1}$=1 nm for (c) T-shaped and (d) PPO configuration. The increase in the number of charged sites $v$ =37, 61, 91 and 469 is represented by dashed-dotted,

dotted, dashed and solid lines. Figures 13 (b), (c) and (d) reproduced with permission from reference 168. Copyright 2004 American Physical Society.

One of the most noteworthy efforts in simulating microstructure of Laponite dispersion has been carried out by Kutter et al.[167] who consider particle as a hexagonal arrangement of homogeneously distributed $v$ discrete charged sites interrelating through a screened Coulomb potential of the Yukawa form. They investigate structure and phase behavior for two model systems. In model A, platelets are modelled as negatively charged sites only; whereas in model B, positively charged edge is considered along with negative face charge. The total surface charge is kept constant in both the models to -700e, which assumes complete dissociation of $Na^+$ ions from the faces of a particle. In Model B a short-range repulsion term is incorporated to avoid the collapse of the particles. A series of molecular-dynamics (MD) simulations have been performed for three values of $v$ (19, 37 and 61) and Debye screening length (0.96, 3.04 and 9.6 nm) as shown in figure 14(a). Runs made for three Laponite concentrations (1%, 3% and 5% by wt.) result in the identification of different phases namely sol, gel and crystal. With increase in Debye length for Model A, enhanced structuring is observed. Model B, on the other hand, being a more convincing representation of Laponite particle, clearly points to clusters of T-shaped configuration at low concentration and house-of-cards structure at higher concentrations (5 wt. %). The authors conclude that illustration of charge using 61 discrete sites is an agreeable approach to the continuous distribution of charge. The modelling approach proposed by Kutter et al.[167] has been employed in several subsequent studies. Harnau et al.[169] adopt model A and apply PRISM theory to the Laponite platelets. The PRISM theory is a simplified form of RISM (Reference Interaction Site Model) with an assumption that all $v$ sites on Laponite particle are equivalent. They consider dispersions of charged circular thin platelets interacting via repulsive Yukawa potential. For the application of linearized Poisson-Boltzmann theory, the effective charge considered is much smaller than the surface charge of particle owing to the counter-ion condensation. The PRISM theory results are in agreement with the experimentally obtained osmotic pressure - $C_L$ results of Mourchid et al.[15] At high $C_S$, however in the absence of salt, an overestimation has been observed. Likewise, the calculated $S(q)$ is also observed to be in agreement with the SANS data of Ramsay and Lindner[170] for 6.5 wt. % dispersion. In order to construct an effective pair potential for discs within the linearized Poisson-Boltzmann theory, Trizac et al.[171] incorporate the anisotropic effect (Yukawa potential) and non-linear effects of counter-ion condensation (charge renormalization). They investigate the consequences of anisotropic potential on phase transition by MC simulations. The interplay between the increase in repulsion and decrease in Debye length with increase in $C_L$ influences the mesoscopic structure. Using the approach of average value of the two-body potential over the angular degrees of freedom, the authors observe a transition to gel state at $C_L = 8$ wt. % which is still significantly higher than observed experimental values.[171]

Odriozola et al.[168] adapt Kutter et al.'s[167] both the model to investigate the colloidal dispersion of Laponite particles by means of Brownian Dynamics (BD) simulations.[168] The authors model Laponite as a bead of 469 spheres out of which only 61 spheres carried charges and others were kept neutral. They incorporate the anisotropic shape effect by integrating the Langevin equations separately for center of mass and orientation. Similar to the previous studies, in order to avoid interpenetration of the platelets, short ranged repulsive potential was considered. The authors investigate a relatively wide concentration range: 2.53 - 38 wt. % for Debye screening lengths of 1 and 3 nm. For Model A at larger Debye length, structured liquids were observed whereas T shaped and house-of-cards configurations were witnessed for Model B. Interestingly, a novel configuration has been reported for Model B, which they term as parallel partially overlapped (PPO) arrangement wherein surface charges of one platelet faces the edge charges of the other as shown in figure 14(b). They observe a minimum in potential energy around $15\,k_BT$ for PPO configuration and a deeper minimum of about $20\,k_BT$ for T-shaped configuration as shown in figure 14(c) and (d). The unique PPO arrangement has been attributed to the presence of neutral sites that prevented interpenetration. This PPO configuration dominates the structure at a concentration greater than 27.8 wt. %, while the T-shaped is more populated at lower concentrations. Furthermore, the reduction in self-diffusion constant with increase in $C_L$ suggests the formation of a gel phase. However, their model does not account for hydrodynamic interactions, which can be important for Laponite suspension.[168] Mossa et al.[172] carry out extensive BD simulation on 8.9 wt. % Laponite dispersion by considering Model A introduced by Kutter et al.[167] The interaction model consists of 61 charged sites but the viscous damping and Brownian forces act on only 3 dynamical sites. In order to account for the anisotropic shape of the particle, they consider two different friction coefficients in perpendicular and parallel direction of symmetry. The total interaction energy between the two rigid platelets is of repulsive Yukawa form with the Debye screening length of 3 nm. Contrary to the nematic phase reported by Mourchid et al.[6] for 8.9 wt. %, Mossa et al.[172] did not discover any evidence of such order in the system. The authors anticipate that unlike spherical particles interacting through Yukawa potential, anisotropy of platelets might lead to a local disorder which could stabilize metastability of the arrested state.[172] Finally, they conclude that with greater time of simulation, dispersion of charged platelets could transform into a Wigner glass.

Inspired by Model B, Jonsson et al.[173] also model Laponite particle as an array of 656 discrete charged spheres with an aspect ratio of 1:24. The authors study the interaction between two Laponite platelets in an electrolyte solution by performing MC simulations. The total interaction has been expressed as the sum of screened Coulomb potential and van der Waals attraction. The authors calculate the free energy of interaction and based on its variation with distance of separation, they describe a phase diagram for Laponite suspension. For moderate value of $C_L$ and $C_S$ (< 5 mM), they observe the interactions to be strongly repulsive leading to a Wigner glass structure. An increase in $C_S$ reduces repulsion resulting in a liquid-like sol phase. With further increase in $C_S$, the system transforms into an attractive gel state, where PPO configuration has minimum free energy. Finally, at sufficiently high $C_S$ (>100 mM), attractive van der Waals interactions prevails and the system of clay particles undergoes precipitation. In a subsequent work, Jonsson and coworkers[174] study dispersion with a wide range of Laponite (0.8 - 35 wt. %) and salt (1 mM – 100 mM) concentrations. A Laponite particle is modelled as a combination of 199 charged spheres having a smaller aspect ratio of 1:15. The authors construct a phase diagram based on the appearance of peaks in radial distribution function. In dilute regime, same behavior as reported previously is observed.[173] At high charge anisotropy (number of edge sites = 48) and intermediate Laponite $\left(5 < C_L < 12.6\right)$ and salt $\left(C_S \leq 10\text{mM}\right)$ concentration, a liquid crystalline Smectic B phase is reported. This crystalline phase being thermodynamically unstable finally transforms in a stable house of card configuration at higher concentrations. However, at intermediate charge anisotropy (number of edge sites = 38) a direct transition from sol to gel phase iss observed without an intermediate crystalline phase, suggesting the structure to be sensitive to the charge anisotropy. Very interestingly, no Wigner glass state has been observed at higher concentrations. Jonsson and coworkers[174] conclude that the Laponite dispersion is stabilized by the repulsive potential in dilute regime; with increase in $C_L$ and $C_S$ the system is primarily attractive owing to the PPO and house of card conformations.

Ruzicka et al.[116, 127] adapt the primitive model of patchy Laponite discs to study the low and high concentration system. Following the representation of Laponite particle given by Kutter et al.,[167] Ruzicka et al.[116, 127] model Laponite as an assembly of 19 rigid disks without any edge charge. They carry MC simulations with 200 discs and at a lower number density $(\rho)$ than the actual number density of Laponite particles $(\rho_L)$ in the system $(\rho \approx 0.4\rho_L)$.[127] The simulated $S(q)$ with Debye screening length of 5 nm and total charge of 70 $e$ results in good agreement with the experimental $S(q)$ for high concentration samples $(2 < C_L < 3)$. They also calculate the theoretical $S(q)$ by considering only the centers of mass of the particles and treating them as spherical points. The theoretical $S(q)$ obtained by the Percus-Yevick approximation in the absence of attractive interactions with a total charge of 60 $e$ and $\rho \approx 0.4\rho_L$ is reported to be similar to the experimental and simulated $S(q)$. Based on these results, the authors conclude that the microstructure of Laponite dispersion is stabilized by repulsive interactions. Ruzicka et al.[116] further conduct numerical simulation on low concentration Laponite dispersions, wherein they primarily focused on electrostatic attraction between the oppositely charged faces and edges. This has been implemented on the particle surface by incorporating three sites on the edge and one at center of each face. Owing to the edge-face interaction, macroscopic networks has been observed at a relatively low concentration $(1 < C_L < 2)$ of Laponite. They propose a numerical phase diagram based on their MC and Gibbs ensemble MC simulation results. With a decrease in temperature, the system evolves from sol to a percolating structure. However, the coexisting densities reported were significantly greater than the experimental values.[4] The authors also simulate $S(q)$ from MC simulations which are in close agreement with their experimental results.[116] Finally, they propose a phase diagram of Laponite dispersion as a function of waiting time. The gel phase observed at Laponite concentration above 1 wt. % is interpreted as equilibrium gel formed by a percolated network of T-shaped bonds. A transition to Wigner glass is reported at a concentration greater than 2 wt. % which is stabilized by electrostatic repulsion. In a subsequent work by Ruzicka and coworkers,[150] $S(q)$ is simulated using MC simulations by considering both repulsive and attractive interactions in the model. The value of Debye length (≈9 nm) is selected in a fashion to complement the experimental results. For the studied concentration of 3 wt. % Laponite dispersion, initially the system is reported to be stabilized by strong electrostatic repulsion known as Wigner glass. However, with a decrease in temperature (analogous to increase in $t_w$), a relative order in the orientation is observed. They do not find evidence of large number of patchy bonds and also the position of the observed peak in $S(q)$ is smaller than the value corresponding to a percolated network. The authors identify this state as a Disconnected House of Card glass. However, the modelling of Laponite particle as a rigid disc of 19 sites (1:5) does not capture the true aspect ratio of a Laponite particle (1:25).[4]

The various simulation studies carried out on this subject suggests that microstructure obtained in the same is very sensitive to the details of a model as well as other aspects associated with Laponite dispersion such as number density, consideration of dissimilar charges on edge and faces, aspect ratio of the particle, assumed value of Debye screening length, etc. Various simulation studies consider realistic values of only some of the above-mentioned parameters, but not all. Consequently, while simulation studies indeed render insight into the structure and dynamics, they still do not give the definite picture of the microstructure.

**Proposed microstructure and the state diagram**

The discussion in the previous section clearly suggests that the literature presents a variety of opinions regarding the microstructure of Laponite dispersion. There is a consensus that dispersion with a concentration of Laponite below 2 wt. %, irrespective of the concentration of salt, is in an attractive gel state. On the other hand, for concentrations above 2 wt. %, the literature is broadly divided into two schools, one that proposes particles to be present in repulsive glass state,[4, 16] while the other that proposes attractive gel state.[3, 5] Many observations in the literature do not give an explicit indication of a particular microstructure, and the results therefore can be interpreted to suit either of the microstructures.[3] After careful analysis of the literature, we believe that the following three observations give a very clear indication of the microstructure of aqueous Laponite dispersion above 2 wt. % concentration to be in the attractive gel state:

1. The first and foremost is the effect of salt, which is known to accelerate the microstructural build-up that eventually renders the dispersion soft solid-like consistency.[31, 126, 175] Let us assume that the system with $C_L$ > 2 wt. % having no salt gets arrested in a repulsive glassy state wherein there is strong electrostatic repulsion among the Laponite particles. Addition of salt to such system is expected to screen the repulsive potential. Therefore, if the final structure is repulsion dominated, formation of such structure must get delayed in time by addition of salt. However, the very opposite observation that structure formation gets accelerated with the addition of salt clearly suggests the presence of attractive gel microstructure.

2. Furthermore, the effect of tetrasodium pyrophosphate that inhibits the gel formation by shielding the positive charge on the edges of Laponite also clearly point out towards the edge – face interaction, which affirms an attractive gel state of Laponite dispersion.

3. The most explicit confirmation of the attractive gel like microstructure is given by the Cryo-TEM images. As evident from figure 8, for concentration below 2 wt. % the cryo-TEM images indeed show individual Laponite particles forming interparticle edge – face bonds.[5] For concentrations above 2 wt. %, as depicted in figure 9, small tactoids of Laponite participate in the network like structure. However, as also shown in figure 9 nowhere during the evolution of Laponite dispersion the microstructure could be repulsive in nature. It is therefore not surprising that the evolving microstructure of Laponite dispersion shows all the characteristic features of sol – gel transition that is well documented in the literature for polymeric gel forming systems.[5]

Based on Cryo-TEM images, rheology and conductivity measurements, a likely scenario of what may be occurring in Laponite dispersion is presented below.

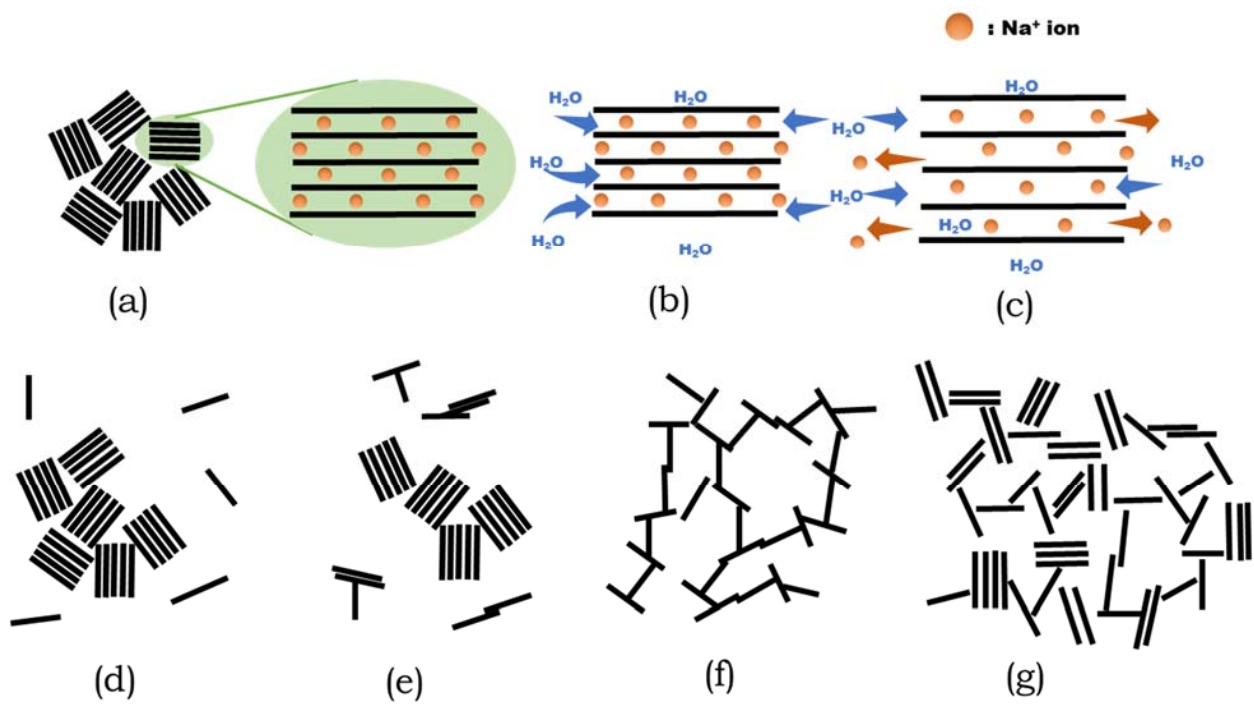

**Figure 15.** Schematic representation of structural evolution in an aqueous dispersion of Laponite (a) In dry form, Laponite is present as aggregates of tactoids. A magnified image of one tactoid is shown. (b) Water enters the interlayer gallery and solvates the $Na^+$ ions present on the faces of the particles. (c) $Na^+$ ions diffuse out of the interlayer gallery thus rendering a negative face charge and the distance between particles increases (d) The delaminated tactoids undergo thermal motion in aqueous medium (e) Gradual delamination of tactoid continues and the particles interact to form clusters (f) Formation of Type 1 gel after prolonged time: A network structure is formed by predominately individual Laponite particles in low concentration Laponite sample ($C_L <$ 2 wt. %, the corresponding cryo-TEM image is given in figure 8) (g) Formation of Type 2 gel: The percolated structure is formed by unexfoliated tactoids of Laponite particles in case of high concentration Laponite dispersion ($C_L > $ 2 wt. %, the corresponding cryo-TEM image is given in figure 9).

In the powder form, Laponite is available in the form of tactoid aggregates with size around 1 μm to 100 μm as represented in schematic figure 15(a). When Laponite powder is dispersed in water, the tactoid aggregates swell and grow with time. Simultaneously the solvated $Na^+$ ions diffuse out of the interlayer gallery causing further swelling due to inter-face repulsion as schematically depicted in figures 15(b) and (c). During this process individual particles or smaller tactoids on the surface of the tactoid aggregates continuously delaminate and diffuse in- bulk as shown in figure 15(d). Therefore, in the initial period, the system has delaminated Laponite particles/small tactoids and swelling aggregates of tactoids. The delaminated particles/small tactoids, that have now acquired positive edge charges and negative face charges, continuously undergo thermal motion. While doing so if they come in contact with each other in such a fashion that their edges and faces interact, they form a bond after overcoming the repulsive energy barrier. As time progresses, the aggregates of tactoids swell and progressively become lean due to delamination of Laponite particles/small tactoids from its periphery. As a result, after an initial period has elapsed there exist original but depleted aggregates of tactoids, individual particles/small tactoids, and newly formed clusters as shown in figure 15(e). The newly formed Laponite clusters grow with time and eventually span the space leading to the formation of a network (gel). For smaller concentrations (typically below 2 wt. %) the space spanning network formation process is so slow that it occurs after the original aggregates of tactoids have almost completely delaminated as represented in figure 15(f) (cryo-TEM image in figure 8). For higher concentrations (above 2 wt. %) space spanning network formation process is fast enough so that it takes place before the delamination of aggregates of tactoids is complete as depicted in figure 15(g) (cryo-TEM image in figure 9). As a result, the depleted aggregates of tactoids remain part of the space spanning structure in the latter case. In such a scenario, the concentration homogenization continues after the space spanning network has formed. In both the cases, the restructuring of the Laponite particles persists to explore progressively low energy states.

We now propose a state diagram for aqueous Laponite dispersion as shown in figure 16. The horizontal axis of the state diagram is the concentration of Laponite while the vertical axis is the concentration of externally added NaCl. It should be noted that the concentration of $Na^+$ ions is usually significantly higher than the concentration of NaCl. Furthermore, the concentration of $Na^+$ ions also increases with time because of dissociation of $Na^+$ counter-ions from the faces of Laponite that continues over a period of at least two weeks.[10] As reported in the literature for $C_L <$ 1 wt. %, the dispersion undergoes phase separation.[3, 15, 123] On the other hand, for dispersions containing salt more than 20 mM flocculation is observed.[3, 15] Above 2.8 wt. % Laponite dispersion has been observed to show nematic order.[6, 117] This concentration threshold has been observed to increase with an increase in $C_S$.[97] The corresponding region has been termed as the nematic gel. In the remaining domain shown in the state diagram, the Laponite dispersion eventually gets arrested in the attractive gel state.[5] Depending upon the concentration of Laponite below and above 2 wt. %, we distinguish between the attractive gel as respectively type 1 and type 2. We believe that at the point of ergodicity breaking type 1 gel is composed more of individual Laponite particles while type 2 is composed more of small tactoids of Laponite particles as shown in the TEM images and also suggested by rheology (distinctly different slopes of $\ln S$ versus $n$ plot below and above 2 wt. %).[5] However, it should be noted that the particles continuously undergo restructuring with time thereby exploring progressively low energy states of the phase space.

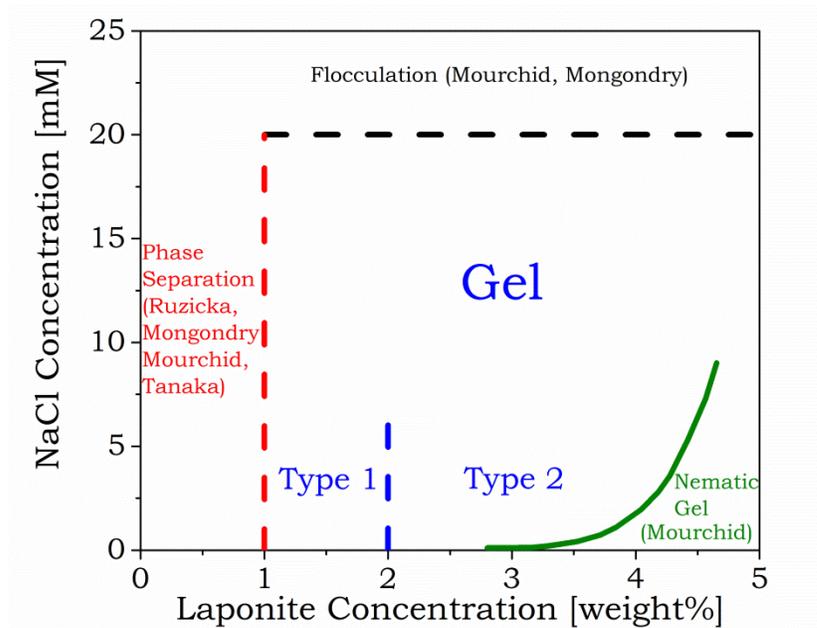

**Figure 16.** The proposed state diagram for aqueous dispersion of Laponite.

**Future Directions and Open Questions**

As discussed in this feature article, a significant amount of work has been carried out on both the fronts: the soft glassy dynamics as well as microstructural investigations of an aqueous dispersion of Laponite. Having said this, there are many questions that are still unanswered on both the fronts. Although both the issues are intertwined, we shall first discuss the open problems in soft glassy dynamics. One of the important concerns is length-scale dependent nature of relaxation dynamics. While the slow relaxation dynamics at the bulk scale is always observed to be stretched exponential,[54] the light scattering results show stretched as well compressed exponential behaviors, which also depend on deformation history of the system.[150] There are disagreements among various groups regarding this observation and indeed deserve further exploration. Furthermore, the effective time domain theory has indeed been observed to lead to validation of various linear viscoelastic principles for the aqueous dispersion of Laponite. However, the transformation of response functions in time domain (such as creep compliance and relaxation modulus) to that of in the frequency domain (such as elastic and viscous moduli) still remains a challenge. Moreover, on one hand, it is difficult to define dynamic moduli for time-dependent materials particularly at low frequencies. On the other hand, contamination of differently aging slow and fast modes may lead to complex dynamic response than that can be predicted using effective time domain theory that assumes shape invariance of relaxation time distribution. Moreover, still there are no constitutive models available that show quantitative predictions of the non-linear response of aqueous dispersion of Laponite. Particularly an important question still is whether dispersion of Laponite shows a true (although time-dependent) yield stress or not. Recently it has been argued that if the evolution of relaxation time with respect to time of material becomes weaker than linear then the material will eventually undergo yielding.[89] It has been observed for Laponite dispersion that the evolution of relaxation time with respect to time does get weaker over time-scale of months.[108] However, it still has not been observed to decrease below unity. This observation is in apparent violation of an important rheological principle: "Everything flows," and the subject needs further attention.[89] Another consequence of stronger than linear relaxation time evolution is non-monotonic steady state – shear rate flow curve, which in turn leads to peculiar phenomena such as time-dependent yield stress, shear banding, and residual stresses. Since decreasing part of the stress – rate flow curve is linearly unstable, how such non-monotonicity of flow curve influences variety of the flow field and how microstructure gets affected by unrelaxed (residual) stresses in a material over a long-term is still open to investigation.

Freshly prepared Laponite dispersion follows Winter-Chambon criterion and this observation is important both from microstructural point of view as well as rheological point of view. The essence of the observation by Winter-Chambon is that the fractal structure is devoid of any specific length-scale and time-scale. As a result, relaxation time distribution shows inverse power law relationship. However, it is still not clear as to how relaxation time distribution evolves to this inverse power law dependence and beyond the same as the Laponite dispersion undergoes critical gelation followed by gel consolidation. The physical meaning of properties such as gel strength ($S$), which has a modulus-scale and time-scale embedded in it, and perceived universal critical exponents are also important open questions that shall further assist in analyzing the microstructure.

Though we have presented a unified picture of microstructure, there are a few subtle aspects that are unexplored or have contrasting observations in the literature. Firstly, it is important to note that Laponite RD/XLG is an industrial product. As a result, it is expected to have some impurities that may include small concentration of different salts. Such aspects, on one hand, may induce errors in calculations of CEC, electrostatic potentials, screening lengths, etc., on the other hand, it will indeed influence the structure, which otherwise is attributed to that of pure Laponite dispersion. Furthermore, it has been observed that point of zero charge of Laponite particle edge is between pH of 10 and 11, the precise number strongly depends on the concentration of Laponite and possibly on the physical state of a dispersion, a subject that deserves further investigation. Some studies have reported significant birefringence near the air-Laponite dispersion interface suggesting strong anisotropic structure.[117] However precise nature of this microstructure, extent to which it affects the gelation in Laponite dispersion, and effect of the presence of various amphiphilic molecules at the interface on the same are still the open questions. Also, the addition of multivalent salts that intricately alters the gelation kinetics cannot be captured by DLVO theory because of strong non-linearities involved in the calculations.[151] It is important to explore this aspect further including how multivalent ions affect the electrostatic interactions between two charged plates. And finally, as we have already discussed, the simulations studies are still far from accounting for the actual system parameters to reflect the realistic microstructural aspects and there is significant scope for improvement.

**Concluding Remarks**

Aqueous dispersion of disk shaped nanoparticles of Laponite, has been a subject of intense investigation over past two decades due to its peculiar out of thermodynamic equilibrium dynamics and industrial importance. After dispersing in water, the particles of Laponite spontaneously self-assemble to form low energy structures that cause evolution of its viscosity and modulus. The consequent increase in the relaxation time with time suggests dynamic slowing down of the constituent's mobility that is reminiscent of physical aging shown by molecular glasses in general and colloidal glasses in particular. As a result, aqueous dispersion of Laponite has been termed as a model soft glassy material that shows various physical as well as rheological characteristic features of the same. Interestingly thixotropic constitutive equation that shows non-monotonic shear stress-shear rate constitutive equation is qualitatively able to capture many of the experimentally observed rheological features. Aqueous dispersion of Laponite has also been observed to validate the fundamental principles of linear viscoelasticity in the effective time domain.

The corresponding evolution of microstructure that transforms liquid into soft solid has also been investigated in the literature using a variety of characterization tools and simulations. While there is a consensus regarding microstructure of dispersion below 2 wt. % to be an attractive gel, various groups debate whether the microstructure above 2 wt. % is composed of attractive gel or a repulsive glass. There are many experimental evidences which either give implicit information about the structure or can be interpreted to validate both the microstructures. Importantly, the cryo-TEM images and the fact that addition of salt, which is expected to reduce the repulsion, accelerates liquid – soft solid transformation process confirm the microstructure of Laponite dispersion to be in the attractive gel state. We suggest a possible mechanism that leads to the formation of two types of gels above and below 2 wt. % concentration. Finally, a new state diagram is proposed that suggests the dominance of attractive interactions while forming the low energy microstructures.

**Acknowledgement:**

We acknowledge financial support from the department of atomic energy – science research council (DAE-SRC), Government of India. We thank Dr. Guruswamy Kumaraswamy, Prof. S. R. Raghavan, Prof. Sanat Kumar, Dr. Ranjini Bandyopadhyay and Dr. Paramesh Gadige for discussion. We also thank Dr. Shweta Jatav for sharing Cryo-TEM images and rheological data with us. The support from Advanced Imaging Center, IIT Kanpur in acquiring the cryo-TEM images is greatly acknowledged.

**References:**

1. R. G. Larson, *The structure and rheology of complex fluids*, Oxford University Press, 1999.
2. J. N. Israelachvili, *Intermolecular and Surfaces Forces*, Academic Press London, 3 edn., 2010.
3. P. Mongondry, J. F. Tassin and T. Nicolai, *Journal of Colloid and Interface Science*, 2005, **283**, 397-405.
4. B. Ruzicka and E. Zaccarelli, *Soft Matter*, 2011, **7**, 1268-1286.
5. S. Jatav and Y. M. Joshi, *Langmuir*, 2017, **33**, 2370-2377.
6. A. Mourchid, E. Lecolier, H. Van Damme and P. Levitz, *Langmuir*, 1998, **14**, 4718-4723.
7. H. Z. Cummins, *Journal of Non-Crystalline Solids*, 2007, **353**, 3891-3905.
8. S. Jatav and Y. M. Joshi, *Journal of Rheology*, 2014, **58**, 1535-1554.
9. H. Van Olphen, *An Introduction to Clay Colloid Chemistry*, Wiley, New York, 1977.
10. A. Shahin and Y. M. Joshi, *Langmuir*, 2012, **28**, 15674-15686.
11. , ed. G. L. Faiza Bergaya, Elsevier, Amsterdam, 2 edn., 2013, vol. 5A.
12. S. L. Tawari, D. L. Koch and C. Cohen, *Journal of Colloid and Interface Science*, 2001, **240**, 54-66.
13. E. C. Bingham, *US Bureau of Standards Bulletin*, 1916, **13**, 309-353.

14. I. Langmuir, *The Journal of Chemical Physics*, 1938, **6**, 873-896.
15. A. Mourchid, A. Delville, J. Lambard, E. Lecolier and P. Levitz, *Langmuir*, 1995, **11**, 1942-1950.
16. S. Jabbari-Farouji, H. Tanaka, G. H. Wegdam and D. Bonn, *Physical Review E*, 2008, **78**, 061405.
17. M. Kroon, W. L. Vos and G. H. Wegdam, *Physical Review E*, 1998, **57**, 1962.
18. T. P. Dhavale, S. Jatav and Y. M. Joshi, *Soft Matter*, 2013, **9**, 7751-7756.
19. S. Morariu and M. Bercea, *The Journal of Physical Chemistry B*, 2011, **116**, 48-54.
20. A. Shahin and Y. M. Joshi, *Physical Review Letters*, 2011, **106**, 038302.
21. A. S. Negi and C. O. Osuji, *Physical Review E*, 2009, **80**, 010404.
22. S. Jatav and Y. M. Joshi, *Faraday discussions*, 2016, **186**, 199-213.
23. A. K. Gaharwar, P. J. Schexnailder, B. P. Kline and G. Schmidt, *Acta Biomaterialia*, 2011, **7**, 568-577.
24. C. Mousty, *Applied Clay Science*, 2004, **27**, 159-177.
25. P. Labbé, B. Brahimi, G. Reverdy, C. Mousty, R. Blankespoor, A. Gautier and C. Degrand, *Journal of Electroanalytical Chemistry*, 1994, **379**, 103-110.
26. J.-L. Besombes, S. Cosnier and P. Labbe, *Talanta*, 1997, **44**, 2209-2215.
27. Z. Navrátilová and P. Kula, *Electroanalysis*, 2003, **15**, 837-846.
28. L. Bippus, M. Jaber and B. Lebeau, *New Journal of Chemistry*, 2009, **33**, 1116-1126.
29. Y. Qi, M. Al-Mukhtar, J. F. Alcover and F. Bergaya, *Applied Clay Science*, 1996, **11**, 185-197.
30. R. Avery and J. Ramsay, *Journal of Colloid and Interface Science*, 1986, **109**, 448-454.
31. L. Li, L. Harnau, S. Rosenfeldt and M. Ballauff, *Physical Review E*, 2005, **72**, 051504.
32. D. W. Thompson and J. T. Butterworth, *Journal of Colloid and Interface Science*, 1992, **151**, 236-243.
33. N. Hegyesi, R. T. Vad and B. Pukánszky, *Applied Clay Science*, 2017, **146**, 50-55.
34. Y. M. Joshi, *The Journal of Chemical Physics*, 2007, **127**, 081102.

35. A. Pek-Ing and L. Yee-Kwong, *Applied Clay Science*, 2015, **107**, 36-45.
36. W. Stumm, *Journal*, 1992.
37. C. Martin, F. Pignon, J.-M. Piau, A. Magnin, P. Lindner and B. Cabane, *Physical Review E*, 2002, **66**, 021401.
38. S. Jatav and Y. M. Joshi, *Applied Clay Science*, 2014, **97–98**, 72-77.
39. J. Kreit, I. Shainberg and A. Herbillon, *Clays and Clay Minerals*, 1982, **30**.
40. A. Mourchid and P. Levitz, *Physical Review E*, 1998, **57**, R4887-R4890.
41. R. P. Mohanty and Y. M. Joshi, *Applied Clay Science*, 2016, **119**, 243-248.
42. R. P. Mohanty, K. Suman and Y. M. Joshi, *Applied Clay Science*, 2017, **138**, 17-24.
43. D. Bonn, P. Coussot, H. Huynh, F. Bertrand and G. Debrégeas, *EPL (Europhysics Letters)*, 2002, **59**, 786.
44. I. M. Hodge, *Science*, 1995, **267**, 1945-1947.
45. L. C. E. Struik, *Physical Aging in Amorphous Polymers and Other Materials*, Elsevier, Houston, 1978.
46. S. M. Fielding, M. E. Cates and P. Sollich, *Soft Matter*, 2009, **5**, 2378-2382.
47. B. Abou, D. Bonn and J. Meunier, *Physical Review E*, 2001, **64**, 021510.
48. S. Jabbari-Farouji, G. H. Wegdam and D. Bonn, *Physical review letters*, 2007, **99**, 065701.
49. F. Schosseler, S. Kaloun, M. Skouri and J. Munch, *Physical Review E*, 2006, **73**, 021401.
50. S. Kaloun, R. Skouri, M. Skouri, J. Munch and F. Schosseler, *Physical Review E*, 2005, **72**, 011403.
51. D. Bonn, H. Tanaka, G. Wegdam, H. Kellay and J. Meunier, *EPL (Europhysics Letters)*, 1999, **45**, 52.
52. B. Ruzicka, L. Zulian and G. Ruocco, *Physical review letters*, 2004, **93**, 258301.
53. D. Saha, Y. M. Joshi and R. Bandyopadhyay, *Soft Matter*, 2014, **10**, 3292-3300.
54. R. Bandyopadhyay, P. H. Mohan and Y. M. Joshi, *Soft Matter*, 2010, **6**, 1462-1466.
55. T. Nicolai and S. Cocard, *Journal of colloid and interface science*, 2001, **244**, 51-57.

56. T. Norisuye, M. Inoue, M. Shibayama, R. Tamaki and Y. Chujo, *Macromolecules*, 2000, **33**, 900-905.

57. M. Bellour, A. Knaebel, J. Harden, F. Lequeux and J.-P. Munch, *Physical review E*, 2003, **67**, 031405.

58. F. Ianni, R. Di Leonardo, S. Gentilini and G. Ruocco, *Physical Review E*, 2007, **75**, 011408.

59. L. Zulian, F. A. de Melo Marques, E. Emilitri, G. Ruocco and B. Ruzicka, *Soft matter*, 2014, **10**, 4513-4521.

60. B. Zheng and S. R. Bhatia, *Colloids and Surfaces A: Physicochemical and Engineering Aspects*, 2017, **520**, 729-735.

61. A. Knaebel, M. Bellour, J.-P. Munch, V. Viasnoff, F. Lequeux and J. Harden, *EPL (Europhysics Letters)*, 2000, **52**, 73.

62. F. A. de Melo Marques, R. Angelini, E. Zaccarelli, B. Farago, B. Ruta, G. Ruocco and B. Ruzicka, *Soft Matter*, 2015, **11**, 466-471.

63. H. Tanaka, S. Jabbari-Farouji, J. Meunier and D. Bonn, *Physical Review E*, 2005, **71**, 021402.

64. R. Bandyopadhyay, D. Liang, H. Yardimci, D. Sessoms, M. Borthwick, S. Mochrie, J. Harden and R. Leheny, *Physical Review Letters*, 2004, **93**, 228302.

65. L. Cipelletti and L. Ramos, *Journal of Physics: Condensed Matter*, 2005, **17**, R253.

66. L. Berthier, *arXiv preprint arXiv:1106.1739*, 2011.

67. R. Böhmer, R. Chamberlin, G. Diezemann, B. Geil, A. Heuer, G. Hinze, S. Kuebler, R. Richert, B. Schiener and H. Sillescu, *Journal of non-crystalline solids*, 1998, **235**, 1-9.

68. H. Sillescu, *Journal of Non-Crystalline Solids*, 1999, **243**, 81-108.

69. E. V. Russell and N. Israeloff, *Nature*, 2000, **408**, 695.

70. S. Jabbari-Farouji, R. Zargar, G. Wegdam and D. Bonn, *Soft Matter*, 2012, **8**, 5507-5512.

71. C. Maggi, R. Di Leonardo, G. Ruocco and J. C. Dyre, *Physical review letters*, 2012, **109**, 097401.

72. P. Gadige, D. Saha, S. K. Behera and R. Bandyopadhyay, *Scientific Reports*, 2017, **7**, 8017.
73. F. K. Oppong, P. Coussot and J. R. de Bruyn, *Physical Review E*, 2008, **78**, 021405.
74. J. P. Rich, G. H. McKinley and P. S. Doyle, *Journal of Rheology*, 2011, **55**, 273-299.
75. M. Pilavtepe, S. Recktenwald, R. Schuhmann, K. Emmerich and N. Willenbacher, *Journal of Rheology*, 2018, **62**, 593-605.
76. O. Dauchot, G. Marty and G. Biroli, *Physical review letters*, 2005, **95**, 265701.
77. W. K. Kegel and A. van Blaaderen, *Science*, 2000, **287**, 290-293.
78. D. Saha, R. Bandyopadhyay and Y. M. Joshi, *Langmuir*, 2015, **31**, 3012-3020.
79. D. Saha, Y. M. Joshi and R. Bandyopadhyay, *The Journal of chemical physics*, 2015, **143**, 214901.
80. C. Toninelli, M. Wyart, L. Berthier, G. Biroli and J.-P. Bouchaud, *Physical Review E*, 2005, **71**, 041505.
81. G. L. Hunter and E. R. Weeks, *Reports on Progress in Physics*, 2012, **75**, 066501.
82. B. Abou and F. Gallet, *Physical Review Letters*, 2004, **93**, 160603.
83. N. Greinert, T. Wood and P. Bartlett, *Physical review letters*, 2006, **97**, 265702.
84. S. Jabbari-Farouji, D. Mizuno, M. Atakhorrami, F. C. MacKintosh, C. F. Schmidt, E. Eiser, G. H. Wegdam and D. Bonn, *Physical review letters*, 2007, **98**, 108302.
85. S. Jabbari-Farouji, D. Mizuno, D. Derks, G. Wegdam, F. MacKintosh, C. Schmidt and D. Bonn, *EPL (Europhysics Letters)*, 2008, **84**, 20006.
86. P. Jop, J. R. Gomez-Solano, A. Petrosyan and S. Ciliberto, *Journal of Statistical Mechanics: Theory and Experiment*, 2009, **2009**, P04012.
87. A. Shahin and Y. M. Joshi, *Langmuir*, 2010, **26**, 4219-4225.
88. M. Kaushal and Y. M. Joshi, *Soft Matter*, 2014, **10**, 1891-1894.
89. Y. M. Joshi and G. Petekidis, *Rheologica Acta*, 2018, 1-29.
90. M. T. Shaw and W. J. MacKnight, *Introduction to polymer viscoelasticity*, John Wiley & Sons, 2005.
91. Y. M. Joshi and G. R. K. Reddy, *Physical Review E*, 2008, **77**, 021501.

92. A. S. Negi and C. O. Osuji, *Journal of Rheology*, 2010, **54**, 943-958.
93. N. Willenbacher, *Journal of Colloid and Interface Science*, 1996, **182**, 501-510.
94. Y. M. Joshi, G. R. K. Reddy, A. L. Kulkarni, N. Kumar and R. P. Chhabra, 2008.
95. P. Coussot, H. Tabuteau, X. Chateau, L. Tocquer and G. Ovarlez, *Journal of Rheology*, 2006, **50**, 975-994.
96. B. Baldewa and Y. M. Joshi, *Soft Matter*, 2012, **8**, 789-796.
97. A. Shukla and Y. M. Joshi, *Chemical Engineering Science*, 2009, **64**, 4668-4674.
98. Y. M. Joshi, A. Shahin and M. E. Cates, *Faraday discussions*, 2012, **158**, 313-324.
99. A. S. Negi and C. O. Osuji, *Physical Review E*, 2010, **82**, 031404.
100. A. Shaukat, A. Sharma and Y. M. Joshi, *Journal of Non-Newtonian Fluid Mechanics*, 2012, **167**, 9-17.
101. J. D. Martin and Y. T. Hu, *Soft Matter*, 2012, **8**, 6940-6949.
102. A. Jain, R. Singh, L. Kushwaha, V. Shankar and Y. M. Joshi, *Journal of Rheology*, 2018, **62**, 1001-1016.
103. T. Gibaud, C. Barentin and S. Manneville, *Physical Review Letters*, 2008, **101**, 258302.
104. S. M. Fielding, P. Sollich and M. E. Cates, *Journal of Rheology*, 2000, **44**, 323-369.
105. M. Kaushal and Y. M. Joshi, *Macromolecules*, 2014, **47**, 8041-8047.
106. Y. M. Joshi, *Rheologica Acta*, 2014, **53**, 477-488.
107. R. Gupta, B. Baldewa and Y. M. Joshi, *Soft Matter*, 2012, **8**, 4171-4176.
108. A. Shahin and Y. M. Joshi, *Langmuir*, 2012, **28**, 5826-5833.
109. A. Shukla and Y. M. Joshi, *Rheologica Acta*, 2017, **56**, 927-940.
110. Y. M. Joshi, *Soft matter*, 2015, **11**, 3198-3214.
111. J. Bosse and S. D. Wilke, *Physical Review Letters*, 1998, **80**, 1260-1263.
112. D. Bonn, H. Kellay, H. Tanaka, G. Wegdam and J. Meunier, *Langmuir*, 1999, **15**, 7534-7536.
113. T. Nicolai and S. Cocard, *Langmuir*, 2000, **16**, 8189-8193.
114. S. Bhatia, J. Barker and A. Mourchid, *Langmuir*, 2003, **19**, 532-535.

115. S. Jabbari-Farouji, H. Tanaka, G. Wegdam and D. Bonn, *Physical Review E*, 2008, **78**, 061405.
116. B. Ruzicka, E. Zaccarelli, L. Zulian, R. Angelini, M. Sztucki, A. Moussaïd, T. Narayanan and F. Sciortino, *Nature materials*, 2011, **10**, 56.
117. A. Shahin, Y. M. Joshi and S. A. Ramakrishna, *Langmuir*, 2011, **27**, 14045-14052.
118. T. Li, A. J. Senesi and B. Lee, *Chemical reviews*, 2016, **116**, 11128-11180.
119. S. Hansen, *Journal of Applied Crystallography*, 2013, **46**, 1008-1016.
120. A. Guiner and G. Fournet, *Small angle scattering of X-rays*, J. Wiley & Sons, , New York, 1955.
121. J. S. Pedersen, *Advances in colloid and interface science*, 1997, **70**, 171-210.
122. D. G. Greene, D. V. Ferraro, A. M. Lenhoff and N. J. Wagner, *Journal of Applied Crystallography*, 2016, **49**, 1734-1739.
123. H. Tanaka, J. Meunier and D. Bonn, *Physical Review E*, 2004, **69**, 031404.
124. A. Mourchid, A. Delville and P. Levitz, *Faraday Discussions*, 1995, **101**, 275-285.
125. P. Levitz, E. Lecolier, A. Mourchid, A. Delville and S. Lyonnard, *EPL (Europhysics Letters)*, 2000, **49**, 672.
126. B. Ruzicka, L. Zulian, R. Angelini, M. Sztucki, A. Moussaïd and G. Ruocco, *Physical Review E*, 2008, **77**, 020402.
127. B. Ruzicka, L. Zulian, E. Zaccarelli, R. Angelini, M. Sztucki, A. Moussaïd and G. Ruocco, *Physical Review Letters*, 2010, **104**, 085701.
128. F. A. d. M. Marques, R. Angelini, G. Ruocco and B. Ruzicka, *The Journal of Physical Chemistry B*, 2017, **121**, 4576-4582.
129. V. Thrithamara Ranganathan and R. Bandyopadhyay, *Colloids and Surfaces A: Physicochemical and Engineering Aspects*, 2017, **522**, 304-309.
130. P. Gadige and R. Bandyopadhyay, *Soft Matter*, 2018.
131. S. Jabbari-Farouji, M. Atakhorrami, D. Mizuno, E. Eiser, G. Wegdam, F. MacKintosh, D. Bonn and C. Schmidt, *Physical Review E*, 2008, **78**, 061402.
132. S. Venkataraman and H. Winter, *Rheologica Acta*, 1990, **29**, 423-432.
133. H. H. Winter and F. Chambon, *Journal of Rheology*, 1986, **30**, 367-382.
134. B.-S. Chiou, S. R. Raghavan and S. A. Khan, *Macromolecules*, 2001, **34**, 4526-4533.

135. T. Fuchs, W. Richtering, W. Burchard, K. Kajiwara and S. Kitamura, *Polymer Gels and Networks*, 1998, **5**, 541-559.
136. P. Matricardi, M. Dentini, V. Crescenzi and S. Ross-Murphy, *Carbohydrate polymers*, 1995, **27**, 215-220.
137. X. Hu, J. Fan and C. Y. Yue, *Journal of applied polymer science*, 2001, **80**, 2437-2445.
138. S. Cocard, J. F. Tassin and T. Nicolai, *Journal of Rheology*, 2000, **44**, 585-594.
139. M. Muthukumar, *Macromolecules*, 1989, **22**, 4656-4658.
140. A. Ponton, A. Bee, D. Talbot and R. Perzynski, *Journal of Physics: Condensed Matter*, 2005, **17**, 821.
141. K.-C. Hung, U.-S. Jeng and S.-h. Hsu, *ACS Macro Letters*, 2015, **4**, 1056-1061.
142. T. Matsumoto, M. Kawai and T. Masuda, *Macromolecules*, 1992, **25**, 5430-5433.
143. H. H. Winter, *Macromolecules*, 2013, **46**, 2425-2432.
144. N. P. Bansal, *Journal of the American Ceramic Society*, 1990, **73**, 2647-2652.
145. Ö. Pekcan and S. Kara, *Polymer*, 2001, **42**, 7411-7417.
146. Y. Bai, C. Xiong, F. Wei, J. Li, Y. Shu and D. Liu, *Energy & Fuels*, 2015, **29**, 447-458.
147. A. Izuka, H. H. Winter and T. Hashimoto, *Macromolecules*, 1992, **25**, 2422-2428.
148. D. Strachan, G. Kalur and S. Raghavan, *Physical Review E*, 2006, **73**, 041509.
149. L. Petit, C. Barentin, J. Colombani, C. Ybert and L. Bocquet, *Langmuir*, 2009, **25**, 12048-12055.
150. R. Angelini, E. Zaccarelli, F. A. D. M. Marques, M. Sztucki, A. Fluerasu, G. Ruocco and B. Ruzicka, *Nature communications*, 2014, **5**, 4049.
151. A. Thuresson, M. Ullner, T. r. Åkesson, C. Labbez and B. Jönsson, *Langmuir*, 2013, **29**, 9216-9223.
152. M. Boström, D. Williams and B. Ninham, *Physical Review Letters*, 2001, **87**, 168103.
153. W. Sun, Y. Yang, T. Wang, H. Huang, X. Liu and Z. Tong, *Journal of colloid and interface science*, 2012, **376**, 76-82.
154. A. Thuresson, M. Segad, T. S. Plivelic and M. Skepö, *The Journal of Physical Chemistry C*, 2017, **121**, 7387-7396.

155. P. Mongondry, T. Nicolai and J.-F. Tassin, *Journal of Colloid and Interface Science*, 2004, **275**, 191-196.
156. H. Takeno and W. Nakamura, *Journal*, 2015.
157. J. Lal and L. Auvray, *Molecular Crystals and Liquid Crystals Science and Technology. Section A. Molecular Crystals and Liquid Crystals*, 2001, **356**, 503-515.
158. A. Nelson and T. Cosgrove, *Langmuir*, 2004, **20**, 2298-2304.
159. E. Loizou, P. Butler, L. Porcar, E. Kesselman, Y. Talmon, A. Dundigalla and G. Schmidt, *Macromolecules*, 2005, **38**, 2047-2049.
160. H. A. Baghdadi, H. Sardinha and S. R. Bhatia, *Journal of Polymer Science Part B: Polymer Physics*, 2005, **43**, 233-240.
161. A. K. Atmuri and S. R. Bhatia, *Langmuir*, 2013, **29**, 3179-3187.
162. L. Zulian, B. Ruzicka and G. Ruocco, *Philosophical Magazine*, 2008, **88**, 4213-4221.
163. A. K. Atmuri, G. A. Peklaris, S. Kishore and S. R. Bhatia, *Soft Matter*, 2012, **8**, 8965-8971.
164. M. Dijkstra, J. Hansen and P. Madden, *Physical review letters*, 1995, **75**, 2236.
165. M. Dijkstra, J.-P. Hansen and P. A. Madden, *Physical Review E*, 1997, **55**, 3044.
166. R. Eppenga and D. Frenkel, *Molecular physics*, 1984, **52**, 1303-1334.
167. S. Kutter, J.-P. Hansen, M. Sprik and E. Boek, *The Journal of Chemical Physics*, 2000, **112**, 311-322.
168. G. Odriozola, M. Romero-Bastida and F. d. J. Guevara-Rodriguez, *Physical Review E*, 2004, **70**, 021405.
169. L. Harnau, D. Costa and J.-P. Hansen, *EPL (Europhysics Letters)*, 2001, **53**, 729.
170. J. Ramsay and P. Lindner, *Journal of the Chemical Society, Faraday Transactions*, 1993, **89**, 4207-4214.
171. E. Trizac, L. Bocquet, R. Agra, J. Weis and M. Aubouy, *Journal of Physics: Condensed Matter*, 2002, **14**, 9339.
172. S. Mossa, C. De Michele and F. Sciortino, *The Journal of chemical physics*, 2007, **126**, 014905.
173. B. Jonsson, C. Labbez and B. Cabane, *Langmuir*, 2008, **24**, 11406-11413.

174. M. Delhorme, B. Jonsson and C. Labbez, *Soft Matter*, 2012, **8**, 9691-9704.
175. T. Nicolai and S. Cocard, *The European Physical Journal E: Soft Matter and Biological Physics*, 2001, **5**, 221-227.